\documentclass[twocolumn,pra,aps,floatfix,footinbib]{revtex4-1}

\usepackage{amsmath}
\usepackage{amssymb}
\usepackage{bigints}
\usepackage{comment}
\usepackage{etoolbox}
\usepackage{feynmf}
\usepackage{graphicx}
\usepackage{lipsum}
\usepackage{tikz}
\usepackage{verbatim}
\usepackage{xcolor}

\usepackage[export]{adjustbox}
\usepackage[colorlinks,urlcolor=blue,citecolor=blue,linkcolor=blue]{hyperref}
\usepackage{cleveref}
\usepackage{pagecolor}
\usepackage[normalem]{ulem}

\usetikzlibrary{shapes}

\newcommand{\0}{\mathbf{0}}
\newcommand{\n}{\mathbf{n}}

\newcommand{\x}{\mathbf{x}}
\newcommand{\y}{\mathbf{y}}
\newcommand{\nn}{\nonumber}
\newcommand{\up}{\uparrow}
\newcommand{\down}{\downarrow}

\newcommand{\del}{\partial}

\newcommand{\vect}[1]{{\mathbf#1}}

\renewcommand{\k}{\mathbf{k}}

\definecolor{red1}{rgb}{5.00,0.10,0.20}
\definecolor{blue1}{rgb}{0.29,0.45,1.00}
\definecolor{green1}{rgb}{0.00,0.69,0.31}
\definecolor{orange1}{rgb}{0.90,0.60,0.10}

\newcommand{\redlineA}{\raisebox{2pt}{\tikz{\draw[-,red1,solid,line width=0.90pt](0,0)--(5mm,0);}}}
\newcommand{\bluelineA}{\raisebox{2pt}{\tikz{\draw[-,blue1,solid,line width=0.90pt](0,0)--(5mm,0);}}}
\newcommand{\greenlineA}{\raisebox{2pt}{\tikz{\draw[-,green1,solid,line width=0.90pt](0,0)--(5mm,0);}}}

\newcommand{\redlineB}{\raisebox{2pt}{\tikz{\draw[-,red1,solid,line width=0.75pt](0,0)--(5mm,0);}}}
\newcommand{\bluelineB}{\raisebox{2pt}{\tikz{\draw[-,blue1,solid,line width=0.75pt](0,0)--(5mm,0);}}}
\newcommand{\greenlineB}{\raisebox{2pt}{\tikz{\draw[-,green1,solid,line width=0.75pt](0,0)--(5mm,0);}}}

\newcommand{\bluelineC}{\raisebox{2pt}{\tikz{\draw[-,blue1,solid,line width=0.65pt](0,0)--(6.50mm,0);}}}
\newcommand{\greenlineC}{\raisebox{2pt}{\tikz{\draw[-,green1,dash pattern=on 5pt off 2pt,line width=0.65pt](0,0)--(6.75mm,0);}}}
\newcommand{\graylineA}{\raisebox{2pt}{\tikz{\draw[-,gray,dash pattern=on 3pt off 2pt,line width=0.65pt](0,0)--(6.40mm,0);}}}
\newcommand{\orangelineA}{\raisebox{2pt}{\tikz{\draw[-,orange1,solid,line width=0.65pt](0,0)--(6.50mm,0);}}}

\newcommand{\bluelineD}{\raisebox{2pt}{\tikz{\draw[-,blue1,solid,line width=0.90pt](0,0)--(6.50mm,0);}}}
\newcommand{\greenlineD}{\raisebox{2pt}{\tikz{\draw[-,green1,dash pattern=on 5pt off 2pt,line width=0.90pt](0,0)--(6.75mm,0);}}}
\newcommand{\graylineB}{\raisebox{2pt}{\tikz{\draw[-,gray,dash pattern=on 3pt off 2pt,line width=0.90pt](0,0)--(6.40mm,0);}}}
\newcommand{\orangelineB}{\raisebox{2pt}{\tikz{\draw[-,orange1,solid,line width=0.90pt](0,0)--(6.50mm,0);}}}

\newcommand{\starA}{\raisebox{0pt}{\tikz{\node[draw,scale=0.65,star,star points=5,star point ratio=0.35,fill=orange,orange,rotate=180](){};}}}
\newcommand{\starB}{\raisebox{0pt}{\tikz{\node[draw,scale=0.55,star,star points=5,star point ratio=0.35,fill=orange,orange,rotate=180](){};}}}

\begin{document}


\footnotetext[1]{Bosonic alkaline-earth atoms do not offer the same utility, here, because their nuclei always have vanishing spins.  The reason is that the even number of electrons (from the full outer shell) implies an even number of protons and, in turn, an even number of neutrons to render the atom a boson.  But for an even-even nucleus, identical nucleons antialign under the influence of the pairing interaction, which therefore leads to an overall nuclear spin of zero~\cite{Scazza_Thesis}.}

\footnotetext[2]{A decay from the excited state, $^{2S+1}L_J=$ $^3P_0$, back to the ground state, $^1S_0$, is doubly forbidden in the dipole approximation, since it would involve a spin flip ($\Delta S\neq0$) and also because we cannot transition between two $J=0$ states.  Consequently, $^3P_0$ is metastable with a lifetime that exceeds $\sim10$ seconds~\cite{PhysRevA.69.021403}.}

\footnotetext[3]{The experiment in Ref.~\cite{PhysRevLett.115.265302} employed a 3D optical lattice at a wavelength of $759.3\,\mathrm{nm}$ and with a depth of 29.8 times the recoil energy, giving a confinement frequency $\omega\simeq2\pi\times20\,\mathrm{kHz}$.}

\footnotetext[4]{We use scattering lengths for the $^{173}$Yb system that have been experimentally determined in Ref.~\cite{PhysRevLett.115.265302}:  $a_S\simeq219.7\,a_0$ and $a_T\simeq1878\,a_0$, where $a_0=0.529\,\text{\AA}$ is the Bohr radius.  For simplicity, we set the singlet and triplet effective ranges, $r_S\simeq126\,a_0$ and $r_T\simeq216\,a_0$, to zero.  We do not expect this simplification to affect the relative shifts between the two- and three-body results in Fig.~\ref{fig:3-Body_Spectrum}, nor the relative shifts between the spin-balanced and\protect\linebreak-imbalanced systems in Fig.~\ref{fig:Many-Body_Plot_(A)}.}

\footnotetext[5]{We see in Fig.~\ref{fig:OFR_Interaction} that the atom pair ($g{\up},\,e{\down}$) has a lower energy than the pair ($g{\down},\,e{\up}$), which leads to the labels of ``open'' and ``closed'' for the scattering channels.  However this designation is somewhat arbitrary since, by interchanging the channel labels and reversing the sign of the detuning $\delta(B)$ (e.g., taking $B\to-B$), the system is left invariant.}

\footnotetext[6]{There is a large decoupling between the electronic-orbital and nuclear-spin degrees of freedom since the total electronic angular momentum ($J$) for both $|g\protect\rangle$ and $|e\protect\rangle$ is zero.  This bestows on the system an SU($N$) symmetry ($N\leq2F+1$)~\cite{Cazalilla_2014} where the collisions do not depend on nuclear spin.  Both groups~\cite{PhysRevLett.115.265302,PhysRevLett.115.265301} who realized the OFR demonstrated this by repeating their measurements for different spin states, and then rescaling the magnetic field to collapse the data onto a single curve.  For our calculations with $^{173}$Yb $\bigl(F=I=\frac52\bigr)$, we assign $m_{F,1}=+\frac52\equiv|{\up}\protect\rangle$ and $m_{F,2}=-\frac52\equiv|{\down}\protect\rangle$.}

\footnotetext[7]{We can see this another way by considering the permutation operator $P_{\up,\down}$ which, when acted on a two-body state, exchanges the nuclear spins.  Now $P_{\up,\down}$ has orbital symmetric and antisymmetric eigenstates, $|g{\down},e{\up}\protect\rangle\mp|g{\up},e{\down}\protect\rangle$ with eigenvalues $\mp1$ (here written in terms of the open and closed channels).  While $P_{\up,\down}$ is only concerned with nuclear spin, the interaction operator $V$ in Eq.~\eqref{eq:Sec2_V(a)} is only concerned with the outer electron densities of the colliding atoms.  Since the nuclear and electronic degrees of freedom in the system are strongly decoupled~\cite{Note6}, we know that $P_{\up,\down}$ and $V$ must have a vanishing commutator and therefore simultaneous eigenstates.}

\footnotetext[8]{The mass of the fermionic isotope $^{173}$Yb is $172.9\,\mathrm{amu}$.}

\footnotetext[9]{For convenience, the factors of $\sqrt{2}$ are chosen so that the effective masses for the center-of-mass motion and the relative motion are the same.}


\title{Frustrated orbital Feshbach resonances in a Fermi gas}

\author{E.~K.~Laird}
\affiliation{School of Physics and Astronomy, Monash University, Victoria 3800, Australia}

\author{Z.-Y.~Shi}
\affiliation{School of Physics and Astronomy, Monash University, Victoria 3800, Australia}

\author{M.~M.~Parish}
\affiliation{School of Physics and Astronomy, Monash University, Victoria 3800, Australia}

\author{J.~Levinsen}
\affiliation{School of Physics and Astronomy, Monash University, Victoria 3800, Australia}

\date{\today}

\begin{abstract}
The orbital Feshbach resonance (OFR) is a novel scheme for magnetically tuning the interactions in closed-shell fermionic atoms.  Remarkably, unlike the Feshbach resonances in alkali atoms, the open and closed channels of the OFR are only very weakly detuned in energy.  This leads to a unique effect whereby a medium in the closed channel can Pauli block --- or frustrate --- the two-body scattering processes.  Here, we theoretically investigate the impact of frustration in the few- and many-body limits of the experimentally accessible three-dimensional $^{173}$Yb system.  We find that by adding a closed-channel atom to the two-body problem, the binding energy of the ground state is significantly suppressed, and by introducing a closed-channel Fermi sea to the many-body problem, we can drive the system towards weaker fermion pairing.  These results are potentially relevant to superconductivity in solid-state multiband materials, as well as to the current and continuing exploration of unconventional Fermi-gas superfluids near the OFR.
\end{abstract}

\pacs{}

\maketitle


\section{Introduction}
\label{sec:Introduction}

Ultracold atomic gases are a testbed for new quantum physics owing to their remarkable tunability.  Factors such as dimensionality, density, and the atoms' internal states and interactions, can all be independently controlled and measured.  This flexibility has been enabled by concurrent experimental advances including the manipulation of optical potentials, radiofrequency spectroscopic techniques and Feshbach resonances~\cite{doi:10.1080/00018730701223200,RevModPhys.80.885,RevModPhys.82.1225}.  In Bose vapours, the achievement of Bose-Einstein condensation~\cite{Anderson198,PhysRevLett.75.1687,PhysRevLett.75.3969} has led notably to observations of the Mott insulator-superfluid phase transition~\cite{PhysRevLett.81.3108,SFMottInsPT_Nat} and bosonic Luttinger liquids~\cite{LuttL_Nat,Kinoshita1125}.  On the other hand, the realization of degeneracy in Fermi gases~\cite{DeMarco1703,Truscott2570,PhysRevLett.87.080403} has allowed for demonstrations of the crossover from a Bardeen-Cooper-Schrieffer (BCS) superfluid of weakly bound Cooper pairs to a molecular Bose-Einstein condensate (BEC)~\cite{PhysRevLett.92.040403,PhysRevLett.92.120401,PhysRevLett.92.120403,PhysRevLett.93.050401}.  Studies of the latter have incorporated systems with unequal populations for the spin-up and spin-down components~\cite{Zwierlein492,Partridge503}, which feature much richer phase diagrams compared to the equal-spin case~\cite{NatPhys_3_124-128_2007,Sheehy2007}.

Feshbach resonances are a tool for controlling the interparticle interactions in a quantum fluid~\cite{RevModPhys.82.1225}.  The basic premise relies on changing the relative energy between two coupled and pairwise potentials that vary differently with an applied magnetic field due to their unequal magnetic moments.  When the threshold of an energetically accessible potential, or open channel, is tuned to match a bound state in an energetically closed channel, the scattering length becomes strongly dependent on the field.  This allows the system to be brought into the unitarity regime of strongest possible interactions.  For alkali gases, the two scattering channels correspond to whether the total spin from the unpaired electrons is a singlet or triplet.  By contrast, and owing to their completely filled outer shells, the total electronic spin of alkaline-earth(-like) atoms in the ground state is zero.  Hence, these species were long thought to be inert against such tuning techniques.

However, in 2015, Zhang et al.~\cite{PhysRevLett.115.135301} proposed a different kind of Feshbach resonance for closed-shell fermions~\cite{Note1} by considering both electronic-orbital and nuclear-spin angular-momentum degrees of freedom.  Their proposal was experimentally feasible since, in addition to the ground state $^1S_0$, the alkaline-earth system features a long-lived first excited state $^3P_0$, where one electron is excited to a $p$ orbital and the two valence electrons form a spin triplet~\cite{Note2}.  In this case, the resonance channels are coupled by so-called interorbital nuclear-spin exchange interactions~\cite{NatPhys_10_779-784_2014,PhysRevLett.113.120402,Zhang1467} which depend on whether the orbital degree of freedom is symmetric or antisymmetric.  Thus, the mechanism was named the orbital Feshbach resonance (OFR)~\cite{PhysRevLett.115.135301}.

Shortly after its prediction, the OFR was experimentally observed in a gas of $^{173}$Yb atoms by two separate groups~\cite{PhysRevLett.115.265302,PhysRevLett.115.265301}.  This has permitted the experimental study of strongly interacting multiorbital Fermi systems, including polarons~\cite{PhysRevLett.122.193604} and ultracold molecules~\cite{PhysRevX.9.011028}.  Concurrently, the OFR has attracted much attention as a promising candidate for realizing a Fermi superfluid in nonalkali atomic gases~\cite{PhysRevLett.115.135301,PhysRevA.94.033609,PhysRevA.94.043624,Mondal_2018,doi:10.7566/JPSJ.87.084302,PhysRevA.95.043634,PhysRevA.93.042708,PhysRevA.95.013618,PhysRevA.95.013624,PhysRevA.95.041603,PhysRevA.96.050701,PhysRevA.97.043616}.  The orbital degree of freedom has furthermore led to progress towards the implementation of the Kondo model in cold atoms~\cite{PhysRevLett.120.143601}.

For the familiar case of alkali resonances, the detuning between the open and closed channels far exceeds all other energy scales in the problem~\cite{RevModPhys.82.1225}.  Consequently, the closed-channel state can be treated as a structureless boson and is only virtually involved in the collisional process.  However, because the nuclear Zeeman effect is much smaller than its electronic counterpart, the channels of the OFR are only very weakly detuned~\cite{PhysRevLett.115.135301,PhysRevA.76.022510}.  As a result, an inert medium in the closed channel can effectively Pauli block --- or frustrate --- atoms scattering from the open channel.  This possibility of frustrating the interactions is truly unique to the OFR but has so far only been considered in the polaron scenario~\cite{Hofer_Thesis,PhysRevLett.122.193604,FrontPhys_Polarons}.

In this work, we theoretically study the impact of frustration on both the few- and the many-body physics near the OFR in three-dimensional $^{173}$Yb.  We first introduce an extra closed-channel atom to the two-body problem and we solve for the new energy spectrum.  Importantly, we observe a sizeable suppression of the ground-state binding energy for the frustrated system, which would not occur if the closed channel was strongly detuned.  Second, we consider the effect of a closed-channel Fermi sea on the mean-field BCS-BEC crossover physics at zero temperature.  Complementing our three-body result, we see that as the Fermi sea is enlarged, both the average chemical potential and the pairing order parameters for the resonance channels move increasingly towards the BCS side of weaker pairing.  We discuss the significance and applications of this new closed-channel physics to current experiments.

This paper is organized as follows:  In Sec.~\ref{sec:Orbital_Feshbach_Resonance}, we introduce our model for studying the orbital Feshbach resonance in $^{173}$Yb.  We conduct our few- and many-body studies on frustration in Secs.~\ref{sec:Three-Body_Problem} and~\ref{sec:Many-Body_Problem}, respectively, and we conclude in Sec.~\ref{sec:Conclusions_and_Outlook}.

\section{Orbital Feshbach Resonance}
\label{sec:Orbital_Feshbach_Resonance}

\begin{figure}[ht]
\begin{center}
\includegraphics[scale=0.3]{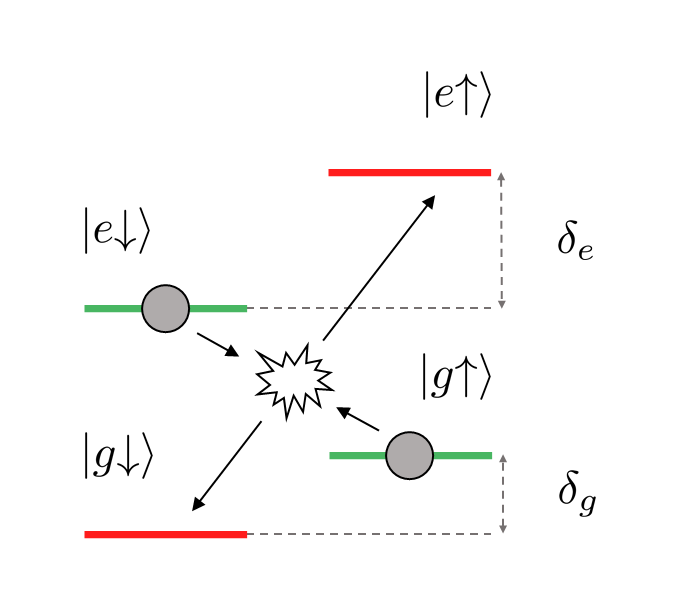}
\caption{A depiction of the OFR pairwise interaction for closed-shell fermions:  The Zeeman energy separations for the ground $|g\rangle$ and excited $|e\rangle$ orbital levels are $\delta_g=g_g(\delta m_F)\mu_BB$ and $\delta_e=g_e(\delta m_F)\mu_BB$, respectively.  Two atoms that begin in the open channel $|g{\up},e{\down}\rangle$ can scatter into the closed channel $|g{\down},e{\up}\rangle$ via an interorbital nuclear-spin exchange process (arrows) which couples the channels together.  At zero magnetic field ($B=0$) the scattering channels are degenerate, but when $B>0$ their energies vary by a small amount, $\delta$ $=\delta_e-\delta_g=(\delta\mu)B$, where $\delta\mu=(\delta g)(\delta m_F)\mu_B$ is the differential magnetic moment.  Here, $\delta g=g_e-g_g$ is the differential Land\'e $g$ factor, $\delta m_F=m_{F,1}-m_{F,2}$ is the difference in quantum number for the nuclear-spin projection, and the Bohr magneton is $\mu_B$~\cite{PhysRevLett.115.135301,PhysRevA.76.022510}.  In $^{173}$Yb, measurements give $\delta\mu/\delta m_F\simeq 2\pi\times110.7\,\mathrm{Hz/G}$~\cite{PhysRevLett.122.193604}.}
\label{fig:OFR_Interaction}
\end{center}
\end{figure}

We describe the orbital Feshbach resonance between a pair of alkaline-earth(-like) fermionic atoms in free space by following Zhang et al.~\cite{PhysRevLett.115.135301} (also see Refs.~\cite{PhysRevA.94.033609} and~\cite{PhysRevA.93.042708}).  First, we take as our degrees of freedom two electronic-orbital states, ``ground'' $^1S_0\equiv|g\rangle$ and ``excited'' $^3P_0\equiv|e\rangle$, and two nuclear-spin states, $m_{F,1}\equiv|{\up}\rangle$ and $m_{F,2}\equiv|{\down}\rangle$~\cite{Note6}.  Then the open channel is designated by one atom in the state $|g{\up}\rangle$ with the other in $|e{\down}\rangle$, while the closed channel has one atom in $|g{\down}\rangle$ and the other in $|e{\up}\rangle$~\cite{Note5}.  We write them as $|g{\up},e{\down}\rangle$ and $|g{\down},e{\up}\rangle$, respectively (and we do not need to antisymmetrize these states since the fermions are distinguishable).  Both channels have the same threshold energies in the absence of a magnetic field $B$;  however, under a finite field they separate by an amount $\delta\equiv\delta(B)$ as shown in Fig.~\ref{fig:OFR_Interaction}.  This effect is attributed to the small difference in nuclear Land\'e $g$ factor between the $|g\rangle$ and $|e\rangle$ orbital levels, which originates from a weak hyperfine coupling of the $^3P_0$ to the $^3P_1$ state~\cite{PhysRevA.76.022510}.  When the interchannel separation $\delta$ is magnetically tuned to match the binding energy of a molecule in the closed channel, the open-channel scattering length diverges, giving rise to the OFR.

Now the noninteracting Hamiltonian, which governs the system when the atoms are far apart, must be diagonal in the open- and closed-channel basis and is given by
\begin{multline} \label{eq:Sec2_H0}
H_0=\sum_{\k}\left[\epsilon_\k\left(g_{\up\k}^\dagger g_{\up\k}+e_{\down\k}^\dagger e_{\down\k}\right)\right.\\+\left.(\epsilon_\k+\delta/2)\left(g_{\down\k}^\dagger g_{\down\k}+e_{\up\k}^\dagger e_{\up\k}\right)\right]\hspace{-0.10em},\tag{1}
\end{multline}
where the single-particle dispersion $\epsilon_\k=|\k|^2/(2m)\equiv k^2/(2m)$ and $m$ is the atomic mass~\cite{Note8}.  Here, $g_{\sigma\k}^{\dagger}$ and $e_{\sigma\k}^{\dagger}$ ($g_{\sigma\k}$ and $e_{\sigma\k}$) are the operators that create (annihilate) a fermion in the ground ($g$) or excited ($e$) electronic orbital state, with three-dimensional (3D) momentum vector $\k$ and nuclear pseudospin $\sigma\in\{\up,\,\down\}$.  Note that, in this work, we employ units where both the system volume and the reduced Planck constant $\hbar$ are set to unity.

When the atoms collide, fermionic statistics imply that the relative wave function is antisymmetric under their exchange.  We know that the spatial component of the wave function is symmetric, because at ultralow temperatures we only have spherical ($s$-wave) scattering.  Correspondingly the internal state, which encompasses both orbital and (nuclear) spin degrees of freedom, must be overall antisymmetric.  This allows for two types of scattering:  one orbital triplet (nuclear-spin singlet) channel and another orbital singlet (nuclear-spin triplet) channel~\cite{Note7}.

One can therefore diagonalize the interaction potential in orbital singlet ($S$) and triplet ($T$) states as follows:
\begin{align} \label{eq:Sec2_V(a)}
V=\sum_{\vect{Q},\,\k,\,\k'}\left(U_SS_{\vect{Q}\k}^\dagger S_{\vect{Q}\k'}+U_TT_{\vect{Q}\k}^\dagger T_{\vect{Q}\k'}\right)\hspace{-0.10em},\tag{2}
\end{align}
with the operators
\begin{align} \label{eq:Sec2_ST}
S_{\vect{Q}\k}^\dagger&=\frac{1}{\sqrt{2}}\left(g_{\down\vect{Q}/2+\k}^\dagger e_{\up\vect{Q}/2-\k}^\dagger+g_{\up\vect{Q}/2+\k}^\dagger e_{\down\vect{Q}/2-\k}^\dagger\right)\hspace{-0.10em},\nn\\T_{\vect{Q}\k}^\dagger&=\frac{1}{\sqrt{2}}\left(g_{\down\vect{Q}/2+\k}^\dagger e_{\up\vect{Q}/2-\k}^\dagger-g_{\up\vect{Q}/2+\k}^\dagger e_{\down\vect{Q}/2-\k}^\dagger\right)\hspace{-0.10em},\tag{3}
\end{align}
and the corresponding interaction strengths, $U_S$ and $U_T$.  This describes the scattering of two particles with center-of-mass momentum $\vect{Q}$, and relative momenta $\vect{k'}$ and $\k$ before and after the collision, respectively.  The full Hamiltonian is then $H=H_0+V$.

By the process of renormalization, the two parameters of the model $\{U_S,\,U_T\}$ can be related to the physical parameters of low-energy collisions, the $s$-wave scattering lengths $\{a_S,\,a_T\}$.  This yields the relations
\begin{align} \label{eq:renormalization}
\frac{m}{4\pi a_{S,\,T}}=\frac{1}{U_{S,\,T}}+\sum_\k^\Lambda\frac{1}{2\epsilon_{\k}}\hspace{+0.15em},\tag{4}
\end{align}
where we set the cutoff $\Lambda$ to infinity at the end of the calculation, while the singlet and triplet scattering lengths~\cite{Note4} are fixed by the properties of the atoms.

When rotated into the same basis as $H_0$, the interaction potential takes the form
\begin{align} \label{eq:Sec2_V(b)}
V=\sum_{\vect{Q},\,\k,\,\k'}\Bigl[U_+\Bigl(&g_{\up\vect{Q}/2+\k}^\dagger e_{\down\vect{Q}/2-\k}^\dagger e_{\down\vect{Q}/2-\k'}g_{\up\vect{Q}/2+\k'}\nn\\+\,\,&g_{\down\vect{Q}/2+\k}^\dagger e_{\up\vect{Q}/2-\k}^\dagger e_{\up\vect{Q}/2-\k'}g_{\down\vect{Q}/2+\k'}\Bigr)\nn\\+\,\,U_-\Bigl(&g_{\up\vect{Q}/2+\k}^\dagger e_{\down\vect{Q}/2-\k}^\dagger e_{\up\vect{Q}/2-\k'}g_{\down\vect{Q}/2+\k'}\nn\\+\,\,&g_{\down\vect{Q}/2+\k}^\dagger e_{\up\vect{Q}/2-\k}^\dagger e_{\down\vect{Q}/2-\k'}g_{\up\vect{Q}/2+\k'}\Bigr)\Bigr]\hspace{+0.05em},\tag{5}
\end{align}
with $U_\pm=(U_S\pm U_T)/2$.  The $U_+$ term denotes an intrachannel potential for both open and closed interactions, while $U_-$ describes an interchannel coupling as illustrated in Fig.~\ref{fig:OFR_Interaction}.

From the Lippmann-Schwinger equation~\cite{PhysRev.79.469}, we now construct the $T$-matrix operator:
\begin{align} \label{eq:T_operator}
T(E)=V+V\frac{1}{E-H_0+i0}\,T(E)\hspace{+0.15em},\tag{6}
\end{align}
where $i0$ denotes an infinitesimal positive imaginary quantity and $E$ the energy.  Importantly, in the center-of-mass frame, the matrix representation of Eq.~\eqref{eq:T_operator} is independent of incoming and outgoing relative momentum.  We can write this conveniently in the singlet and triplet basis,
\begin{align} \label{eq:T_matrix_(A)}
\mathbb{T}(E)=\mathbb{V}+\mathbb{V}\mathbb{R}\sum_\k\begin{pmatrix}\frac{1}{E-2\epsilon_\k+i0}&0\\0&\frac{1}{E-2\epsilon_\k-\delta+i0}\end{pmatrix}\mathbb{R}\mathbb{T}(E)\hspace{+0.15em},\tag{7}
\end{align}
where
\begin{align} \label{eq:V_and_R_matrices}
\mathbb{V}=\left(\begin{array}{cc}U_S&0\\0&U_T\end{array}\right)\quad\mathrm{and}\quad\mathbb{R}=\frac{1}{\sqrt{2}}\left(\begin{array}{cc}1&1\\1&-1\end{array}\right)\tag{8}
\end{align}
are, respectively, the interaction matrix and the involutory matrix that transforms between the singlet-triplet and open-closed channel bases.

Equation~\eqref{eq:T_matrix_(A)} may be readily solved to give
\begin{align} \label{eq:T_matrix_(B)}
\mathbb{T}(E)=\left[\mathbb{V}^{-1}-\mathbb{R}\sum_\k\begin{pmatrix}\frac{1}{E-2\epsilon_\k+i0}&0\\0&\frac{1}{E-2\epsilon_\k-\delta+i0}\end{pmatrix}\mathbb{R}\right]^{-1}\hspace{-0.15em}.\tag{9}
\end{align}
The interactions are then renormalized according to the prescription in Eq.~\eqref{eq:renormalization}, yielding
\begin{align} \label{eq:T_matrix_(C)}
&\mathbb{T}(E)=\nn\\&\frac{4\pi}{m}\left[\begin{pmatrix}a_S&0\\0&a_T\end{pmatrix}^{\hspace{-0.40em}-1}\hspace{-0.30em}-\mathbb{R}\begin{pmatrix}\sqrt{-mE}&0\\0&\sqrt{m(\delta-E)}\end{pmatrix}\mathbb{R}\right]^{-1}\hspace{-0.15em},\tag{10}
\end{align}
where we have used the fact that
\begin{align} \label{eq:integral}
\sum_\k\left(\frac{1}{E-2\epsilon_\k+i0}+\frac{1}{2\epsilon_\k}\right)=\frac{m}{4\pi}\sqrt{-mE}\hspace{+0.15em}.\tag{11}
\end{align}

To obtain the scattering amplitude $f$ relevant to experiment, we project the $T$ matrix onto the open channel:
\begin{widetext}
\begin{align} \label{eq:scattering_amplitude}
f(E)=-\frac{m}{8\pi}\left(\begin{array}{cc}1&1\end{array}\right)\mathbb{T}(E)\left(\begin{array}{c}1\\1\end{array}\right)=\frac{a_S+a_T-2a_Sa_T\sqrt{m(\delta-E)}}{(a_S+a_T)\bigl[\sqrt{-mE}+\sqrt{m(\delta-E)}\,\bigr]-2a_Sa_T\sqrt{-mE}\sqrt{m(\delta-E)}-2}\hspace{+0.15em},\tag{12}
\end{align}
\end{widetext}
and this immediately affords the open-channel $s$-wave (zero-energy) scattering length,
\begin{align} \label{eq:scattering_length}
a=-f(0)=\frac{2a_Sa_T\sqrt{m\delta}-a_S-a_T}{(a_S+a_T)\sqrt{m\delta}-2}\hspace{+0.15em}.\tag{13}
\end{align}
Here, we see that $a$ diverges at $\delta=4/\bigl[m(a_S+a_T)^2\bigr]\equiv\delta_{\mathrm{res}}$, which denotes the magnetic-field position of the OFR.

The pole of the $T$ matrix, Eq.~\eqref{eq:T_matrix_(C)}, provides the condition for a two-body bound state with binding energy $\varepsilon_b$:
\begin{align} \label{eq:pole_condition}
\mathrm{det}\bigl[\mathbb{T}(-\varepsilon_b)\bigr]^{-1}=0\hspace{+0.20em}.\tag{14}
\end{align}
Note, this expression is equivalent to requiring that the open-channel scattering amplitude, Eq.~\eqref{eq:scattering_amplitude}, diverges.  We thus find two molecular states associated with the two fixed scattering lengths.  At zero detuning these have energies $\varepsilon_b=1/\bigl(ma_S^2\bigr)$ and $\varepsilon_b=1/\bigl(ma_T^2\bigr)$.

For most alkaline-earth atoms, the difference in nuclear magnetic moment between the OFR channels, which allows them to be tuned, is very small compared to the alkali case.  Fortunately, in $^{173}$Yb, $a_T$ is unusually large and provides a shallow bound state that can be accessed with magnetic field strengths typical to experiments~\cite{PhysRevLett.113.120402,NatPhys_10_779-784_2014}.  Conversely, $^{173}$Yb also features an extremely small value of $a_S>0$ that gives rise to a very deeply bound state within our model.  It is worth emphasizing that the latter is inaccessible by experiments and not responsible for the OFR.  Moreover, because we are considering low-energy physics, our model cannot properly treat the physics associated with this deeper state.  Thus, we ignore it throughout our paper.

\section{Three-Body Problem}
\label{sec:Three-Body_Problem}

\begin{figure}[t]
\begin{center}
\vspace{2.60em}
\begin{fmffile}{Frustration}
\begin{fmfgraph*}(120,80)
\fmfstraight
\fmfleft{l3,l2,l1}
\fmfright{r3,r2,r1}
\fmf{fermion}{l1,v1,r1}
\fmf{fermion}{l2,v2,r2}
\fmf{fermion}{l3,r3}
\fmf{photon,tension=0}{v1,v2}
\fmflabel{$g{\up}\,,\,\vect{n_1}$}{l1}
\fmflabel{$e{\down}\,,\,\vect{n_2}$}{l2}
\fmflabel{$g{\down}\,,\,\vect{n_3}$}{l3}
\fmflabel{$g{\down}\,,\,\vect{n_1'}\neq\vect{n_3}$}{r1}
\fmflabel{$e{\up}\,,\,\vect{n_2'}$}{r2}
\fmflabel{$g{\down}\,,\,\vect{n_3}$}{r3}
\end{fmfgraph*}
\end{fmffile}
\vspace{2.20em}
\caption{We have two open-channel atoms ($g{\up},\,e{\down}$) and one closed-channel atom ($g{\down}$) in the harmonic-oscillator levels $\vect{n_1},\,\vect{n_2},\,\vect{n_3}$.  After a collision, as drawn, particle 1 is Pauli blocked from occupying the same state as particle 3.}
\label{fig:Feynman_Diagram}
\end{center}
\end{figure}
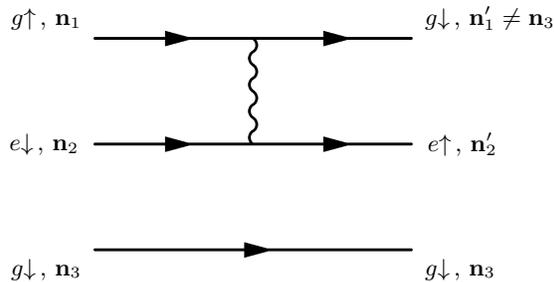

As a first study of frustrated orbital Feshbach resonances, we consider the scenario where two fermionic $^{173}$Yb atoms start out in either the open channel ($g{\up},\,e{\down}$) or the closed channel ($g{\down},\,e{\up}$) and we add a third atom to the closed channel (say $g{\down}$).  We allow the particles to be tightly confined by a three-dimensional and isotropic harmonic-oscillator potential, with frequency $\omega$~\cite{Note3}, which introduces an effective interparticle spacing set by the harmonic-oscillator length, $\sqrt{1/m\omega}$.  Such a model can mimic, in effect, a many-body system where the physics is governed by the scale of the interparticle separation.  Now when two of the atoms, with 3D harmonic-oscillator indices $\vect{n_1}$ and $\vect{n_2}$, scatter from the open channel into the closed channel, the extra atom restricts the number of harmonic-oscillator levels into which they can scatter (see Fig.~\ref{fig:Feynman_Diagram}).  Here, we investigate this Pauli-blocking effect on two-body collisions for the full range of interaction strengths measured in Ref.~\cite{PhysRevLett.115.265302}.

We start from the most general three-body state,
\begin{multline} \label{eq:wave_function}
|\Psi\rangle=\sum_{\vect{n_1},\,\vect{n_2},\,\vect{n_3}}\bigl(\lambda_{\vect{n_1}\vect{n_2}\vect{n_3}}|g_{{\down}\vect{n_1}},e_{{\down}\vect{n_2}},g_{{\up}\vect{n_3}}\rangle\bigr.\\\bigl.+\,\,\tfrac12\gamma_{\vect{n_1}\vect{n_2}\vect{n_3}}|g_{{\down}\vect{n_1}},e_{{\up}\vect{n_2}},g_{{\down}\vect{n_3}}\rangle\bigr)\hspace{+0.10em},\tag{15}
\end{multline}
with basis states $|g_{{\down}\vect{n_1}},e_{{\down}\vect{n_2}},g_{{\up}\vect{n_3}}\rangle\equiv g_{{\down}\vect{n_1}}^\dagger e_{{\down}\vect{n_2}}^\dagger g_{{\up}\vect{n_3}}^\dagger|0\rangle$ and $|g_{{\down}\vect{n_1}},e_{{\up}\vect{n_2}},g_{{\down}\vect{n_3}}\rangle\equiv g_{{\down}\vect{n_1}}^\dagger e_{{\up}\vect{n_2}}^\dagger g_{{\down}\vect{n_3}}^\dagger|0\rangle$, where now the subscripts on the creation operators indicate the harmonic-oscillator indices, and $|0\rangle$ is the vacuum.  Explicitly, $\vect{n_{i}}\equiv\{\rho_{i},\,l_{i},\,\varsigma_{i}\}$ comprises the radial $(\rho=0,\,1,\,...,\,\infty)$, angular-momentum $(l=0,\,1,\,...,\,\rho-1)$, and projection $(\varsigma=-l,\,-l+1,\,...,\,l-1,\,l)$ harmonic-oscillator quantum numbers for the atom at position $\x_i$.  The objects, $\lambda_{\vect{n_1}\vect{n_2}\vect{n_3}}=\langle g_{{\down}\vect{n_1}},e_{{\down}\vect{n_2}},g_{{\up}\vect{n_3}}|\Psi\rangle$ and $\gamma_{\vect{n_1}\vect{n_2}\vect{n_3}}=\langle g_{{\down}\vect{n_1}},e_{{\up}\vect{n_2}},g_{{\down}\vect{n_3}}|\Psi\rangle$, are the wave-function amplitudes and Fermi statistics require $\gamma_{\vect{n_1}\vect{n_2}\vect{n_3}}=-\,\gamma_{\vect{n_3}\vect{n_2}\vect{n_1}}$.

In the presence of the confining potential, the non-interacting Hamiltonian in Eq.~\eqref{eq:Sec2_H0} is modified to
\begin{multline} \label{eq:noninteracting_Hamiltonian}
H_0=\sum_{\vect{n}}\left[\varepsilon_{\vect{n}}\left(g_{{\up}\vect{n}}^\dagger g_{{\up}\vect{n}}+e_{{\down}\vect{n}}^\dagger e_{{\down}\vect{n}}\right)\right.\\\left.+\,\,(\varepsilon_{\vect{n}}+\delta/2)\left(g_{{\down}\vect{n}}^\dagger g_{{\down}\vect{n}}+e_{{\up}\vect{n}}^\dagger e_{{\up}\vect{n}}\right)\right]\hspace{-0.10em}.\tag{16}
\end{multline}
Here, $\varepsilon_{\vect{n}}=(2\rho+l)\omega$ is the single-particle energy eigenvalue for the harmonic oscillator (and we ignore the energy of $\frac32\omega$ corresponding to zero-point motion).

\begin{figure*}[t]
\begin{center}
\includegraphics[scale=0.5]{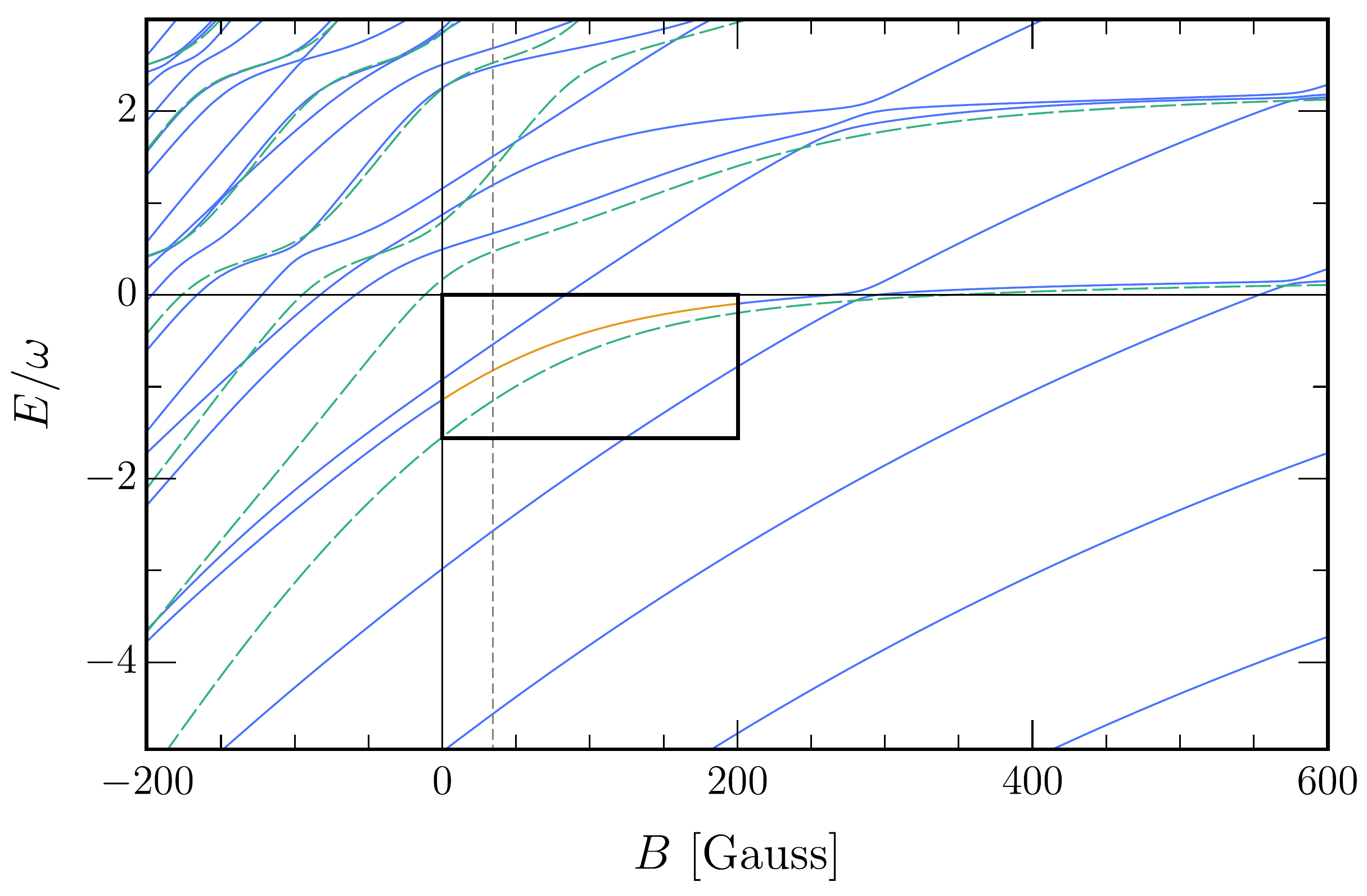}
\caption{Energy spectrum for the frustrated few-body system using ultracold $^{173}$Yb:  The solution to the two-atom problem was first determined in Ref.~\cite{PhysRevLett.115.265302} and corresponds to the green dashed lines (\protect\greenlineC) above.  The blue solid lines (\protect\bluelineC) denote the energies when a third atom is added to the closed channel, as found by numerically solving Eq.~\eqref{eq:matrix_equation}.  We focus on the ground-state shift, in orange (\protect\orangelineA) and highlighted by the box, which is due to Pauli blocking in the closed channel.  (Note that all energies are shown relative to the values for a noninteracting system.)  For reference, the dashed gray vertical line (\protect\graylineA) demarks the orbital Feshbach resonance at $\sim34\,\mathrm{G}$~\cite{Note4}.}
\label{fig:3-Body_Spectrum}
\end{center}
\end{figure*}

The short-range pairwise interactions of the system are diagonal in the basis where the orbital degree of freedom forms a singlet or triplet~\cite{PhysRevLett.115.135301}.  Since they depend only on the relative motion, we describe them by transforming from the coordinates of two colliding atoms, $\vect{x_i}$ and $\vect{x_j}$, to those of their center-of-mass and relative positions, $\vect{X}=(\vect{x_i}+\vect{x_j})/\sqrt{2}$ and $\vect{x_r}=(\vect{x_j}-\vect{x_i})/\sqrt{2}$, respectively~\cite{Note9}.  These have the corresponding harmonic-oscillator indices $\vect{N}$ and $\vect{n_r}$.  We then recast the interacting part, Eq.~\eqref{eq:Sec2_V(a)}, of the Hamiltonian as
\begin{align} \label{eq:interaction_operator}
V=\sum_{\vect{N},\,\vect{n_r},\,\vect{n_r'}}\left(U_SS_{\vect{N}\vect{n_r}}^\dagger S_{\vect{N}\vect{n_r'}}+U_TT_{\vect{N}\vect{n_r}}^\dagger T_{\vect{N}\vect{n_r'}}\right)\hspace{-0.10em},\tag{17}
\end{align}
with the singlet ($S$) and triplet ($T$) operators
\begin{align} \label{eq:singlet_and_triplet_states}
S_{\vect{N}\vect{n_r}}^\dagger&=\frac{\phi_{\rho_r}}{\sqrt{2}}\sum_{\vect{n_1},\,\vect{n_2}}\langle\vect{n_1},\vect{n_2}|\vect{N},\vect{n_r}\rangle\left(g_{{\down}\vect{n_1}}^\dagger e_{{\up}\vect{n_2}}^\dagger+g_{{\up}\vect{n_1}}^\dagger e_{{\down}\vect{n_2}}^\dagger\right)\hspace{-0.10em},\nn\\T_{\vect{N}\vect{n_r}}^\dagger&=\frac{\phi_{\rho_r}}{\sqrt{2}}\sum_{\vect{n_1},\,\vect{n_2}}\langle\vect{n_1},\vect{n_2}|\vect{N},\vect{n_r}\rangle\left(g_{{\down}\vect{n_1}}^\dagger e_{{\up}\vect{n_2}}^\dagger-g_{{\up}\vect{n_1}}^\dagger e_{{\down}\vect{n_2}}^\dagger\right)\hspace{-0.10em}.\tag{18}
\end{align}
At low energy, we only have scattering in the $s$-wave ($l_r=\varsigma_r=0$) because the centrifugal barrier for higher partial waves prevents particles from approaching each other at short distance.  Consequently, $\phi_{\rho_r}$ above denotes the relative $s$-wave harmonic-oscillator eigenfunction at zero separation (see Appendix~\ref{sec:Appendix_A}).  The mentioned coordinate transformation is performed via the two-body Clebsch-Gordan coefficient $\langle\vect{n_1},\vect{n_2}|\vect{N},\vect{n_r}\rangle$, which has the selection rule $\vect{n_1}+\vect{n_2}=\vect{N}+\vect{n_r}$~\cite{SMIRNOV1962346}.

Now that we have set up the problem, we relegate the details of its solution to Appendix~\ref{sec:Appendix_A} and only present the final equations here.  By working in the center-of-mass frame and specializing to the sector of zero total angular momentum, we find that the energy $E$ of the system satisfies
\begin{align} \label{eq:matrix_equation}
\mathrm{det}\left[\mathbb{R}\begin{pmatrix}a_S&0\\0&a_T\end{pmatrix}^{\hspace{-0.40em}-1}\hspace{-0.30em}\mathbb{R}-\begin{pmatrix}\mathbb{A}_E&0\\0&\mathbb{A}_{E-\delta}-\mathbb{B}_{E-\delta}\end{pmatrix}\right]=0\hspace{+0.20em},\tag{19}
\end{align}
where $\{\mathbb{A}_E,\,\mathbb{B}_E\}$ are square matrices given by
\begin{align} \label{eq:matrix_kernels}
\mathbb{A}_E^{\rho,\,\rho'}\hspace{-0.10em}&=\sum_{\kappa}\frac{\sqrt{2}\,\pi \phi_\kappa^2}{E-2(\kappa+\rho)\omega}\delta_{\rho,\,\rho'}\nn\\&=\sqrt{2}\frac{\Gamma[(\rho-E/2)/\omega]}{\Gamma[(\rho-E/2-1/2)/\omega]}\delta_{\rho,\,\rho'}\hspace{+0.15em},\nn\\\mathbb{B}_E^{\rho,\,\rho'}\hspace{-0.10em}&=\sum_{\kappa,\,\kappa'}\frac{\sqrt{2}\,\pi \phi_\kappa \phi_{\kappa'}}{E-2(\kappa'+\rho')\omega}\langle\kappa,\rho|\kappa',\rho'\rangle\hspace{+0.15em},\tag{20}
\end{align}
and $\mathbb{R}$ is the rotation matrix of Eq.~\eqref{eq:V_and_R_matrices}.  The summation indices $\{\kappa,\,\kappa'\}$ refer to the radial quantum numbers for relative atom-atom motion between the interacting particles, while the matrix indices $\{\rho,\,\rho'\}$ are the radial quantum numbers for relative atom-pair motion between the interacting particles' center of mass and the noninteracting atom.  Equation~\eqref{eq:matrix_equation} describes both open-channel scattering (through the two-body term $\mathbb{A}_E$) and closed-channel scattering ($\mathbb{A}_{E-\delta}-\mathbb{B}_{E-\delta}$).  When all three atoms are in the closed channel, the diagonal matrix $\mathbb{A}_{E-\delta}$ is concerned with one of the two interacting pairs.  The nondiagonal kernel $\mathbb{B}_{E-\delta}$ contains the three-body Clebsch-Gordan coefficient $\langle\kappa,\rho|\kappa',\rho'\rangle$ (with $\kappa+\rho=\kappa'+\rho'$), which then allows us to rotate our relative coordinate axes from one interacting pair to the other.  Pauli blocking from the closed channel is manifested by the minus sign in $\mathbb{A}_{E-\delta}-\mathbb{B}_{E-\delta}$, which follows from the symmetry property of the amplitudes, $\gamma_{\vect{n_1}\vect{n_2}\vect{n_3}}=-\,\gamma_{\vect{n_3}\vect{n_2}\vect{n_1}}$, in Eq.~\eqref{eq:wave_function}.  The matrix elements of $\mathbb{A}_E$ in Eq.~\eqref{eq:matrix_kernels} were first treated by Ref.~\cite{Busch_Paper}, while a method for the efficient evaluation of $\mathbb{B}_E$ is provided in Appendix~\ref{sec:Appendix_A}.

Equations~\eqref{eq:matrix_equation}-\eqref{eq:matrix_kernels} can be solved numerically efficiently and the resulting energy spectrum is shown by the solid blue lines (\protect\bluelineD) in Fig.~\ref{fig:3-Body_Spectrum}.  To provide a comparison, the dashed green lines (\protect\greenlineD) give the energies for two atoms interacting in a 3D harmonic trap, which were first calculated by Ref.~\cite{PhysRevLett.115.265302}.  The deep equidistant bound states in the three-body problem arise from the small and positive singlet scattering length of $^{173}$Yb, and are not relevant to the physics of interest, as explained in Sec.~\ref{sec:Orbital_Feshbach_Resonance}.  Instead, we concentrate on the ground state in the region of low-energy physics (orange line, \protect\orangelineB, and highlighted by the box).  Significantly, we see that the binding energy of the ground state is strongly suppressed due to closed-channel Pauli blocking --- by $\sim0.3\,\omega$ at the OFR (gray dashed line, \protect\graylineB).  This physics could, in principle, be probed with clock spectroscopy in a deep optical lattice, similar to the experiment that realized the OFR~\cite{PhysRevLett.115.265302}.  The effects of frustration are stronger and may therefore be easier to observe in this scenario than in free space, where an observation was previously attempted~\cite{PhysRevLett.122.193604}.  However, it remains to be seen whether one would encounter difficulties stemming from three-body losses~\cite{PhysRevX.6.021030}.

\section{Many-Body Problem}
\label{sec:Many-Body_Problem}

\subsection{Theory}
\label{sec:Theory}

Now that we have demonstrated how the few-body correlations can be tailored by adding an atom to the closed channel, we consider how frustration manifests at the many-body level.  To this end, we modify the standard theory~\cite{RevModPhys.80.1215} for describing the zero-temperature BCS-BEC crossover to accommodate a single Fermi sea (say $g{\down}$) in the closed channel near an OFR.  Intuitively, this means that atoms can no longer scatter into the closed channel at an arbitrary momentum, but can only do so within a restricted range of momenta which is set by the Fermi sea.  We stress that the case we consider is different from the more standard scenario of a spin imbalance in the open channel~\cite{PhysRevA.94.053627,PhysRevA.97.013635,EurPhysJ_Polarons}.

The Hamiltonian, $\mathcal{H}=\mathcal{H}_0+V$, for this system consists of the noninteracting component,
\begin{align} \label{eq:many-body_Hamiltonian}
\mathcal{H}_0=&\sum_{\k}\Bigl[\xi_{\k}^{g\up}g_{\up\k}^{\dagger}g_{\up\k}+\xi_{\k}^{e\down}e_{\down\k}^{\dagger}e_{\down\k}+\Bigl(\xi_{\k}^{g\down}+\delta/2\Bigr)g_{\down\k}^{\dagger}g_{\down\k}\nn\\&+\Bigl(\xi_{\k}^{e\up}+\delta/2\Bigr)e_{\up\k}^{\dagger}e_{\up\k}\Bigr]\hspace{+0.05em},\tag{21}
\end{align}
and the interaction potential, $V$ in Eq.~\eqref{eq:Sec2_V(b)}.  Above, $\xi_{\k}^{\alpha}=\epsilon_{\k}-\mu_{\alpha}$ is the single-particle kinetic energy measured relative to the chemical potential $\mu_{\alpha}$, while the fermionic operators are explained below Eq.~\eqref{eq:Sec2_H0}.  Since we have an interconversion process (see Fig.~\ref{fig:OFR_Interaction}) between the open and closed channels, they must be in chemical equilibrium:  $\mu_{g\up}+\mu_{e\down}=\mu_{g\down}+\mu_{e\up}$.  However, this does not place any restriction on the differences, $\mu_{g\up}-\mu_{e\down}$ and $\mu_{g\down}-\mu_{e\up}$.  To allow for the possibility of a closed-channel Fermi sea, we therefore define the average chemical potential $\mu$ and the `Zeeman' field $h$ such that
\begin{alignat}{2} \label{eq:chemical_potentials}
\mu_{g\up}&=\mu\hspace{+0.20em},\qquad\mu_{g\down}&&=\mu+h\hspace{+0.20em},\nn\\\mu_{e\down}&=\mu\hspace{+0.20em},\qquad\mu_{e\up}&&=\mu-h\hspace{+0.20em}.\tag{22}
\end{alignat}

Similar to the approach in Ref.~\cite{PhysRevLett.115.135301}, we now define the real quantities,
\begin{align} \label{eq:gaps}
\widetilde{\Delta}_{o}&=\sum_{\k}\Big\langle e_{\down-\k}g_{\up\k}\Big\rangle\hspace{+0.075em},\nn\\\widetilde{\Delta}_{c}&=\sum_{\k}\Big\langle e_{\up-\k}g_{\down\k}\Big\rangle\hspace{+0.075em},\tag{23}
\end{align}
and we perform a mean-field decoupling of the interaction term, $V$.  This leads to the following reduced Hamiltonian:
\begin{align} \label{eq:mean-field_Hamiltonian_(A)}
&\mathcal{H}^\prime=\frac{U_+\left(\Delta_o^2+\Delta_c^2\right)-2U_-\Delta_o\Delta_c}{U_-^2-U_+^2}+\sum_{\k}\left(2\xi_{\k}+\delta/2+h\right)\nn\\&\hspace{-0.25em}+\sum_{\k}\psi_\k^\dagger\hspace{-0.15em}\left(\begin{array}{cccc}\xi_{\k}&\Delta_o&0&0\\\Delta_o&-\xi_{\k}&0&0\\0&0&\xi_{\k}+\delta/2-h&\Delta_c\\0&0&\Delta_c&-\xi_{\k}-\delta/2-h\\\end{array}\right)\hspace{-0.15em}\psi_\k\hspace{+0.15em},\tag{24}
\end{align}
with $\psi_\k^\dagger=\bigl(g_{\up\k},\,e_{\down-\k}^{\dagger},\,g_{\down\k},\,e_{\up-\k}^{\dagger}\bigr)$ and $\xi_{\k}=\epsilon_{\k}-\mu$.  Here, we have defined the pairing order parameters for the open ($o$) and closed ($c$) channels, $\Delta_o=U_+\widetilde{\Delta}_{o}+U_-\widetilde{\Delta}_{c}$ and $\Delta_c=U_-\widetilde{\Delta}_{o}+U_+\widetilde{\Delta}_{c}$.  Notice that although $\Delta_o$ and $\Delta_c$ of Eq.~\eqref{eq:mean-field_Hamiltonian_(A)} appear to be completely decoupled, they are in fact coupled by the relations in Eq.~\eqref{eq:gaps}.

\begin{figure}[b]
\begin{center}
\includegraphics[scale=0.3]{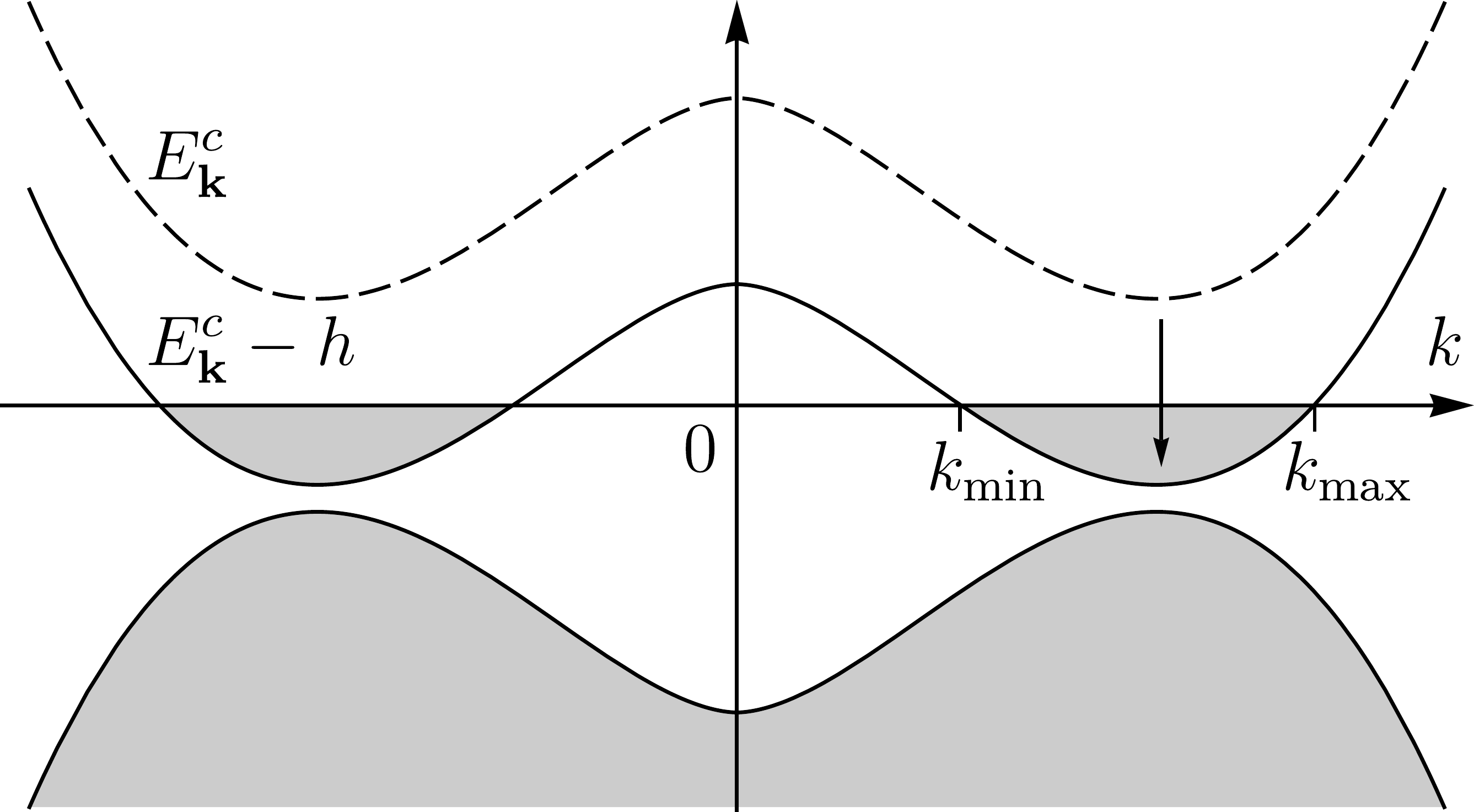}
\caption{Closed-channel excitation spectrum with empty (unshaded) and filled bands (gray shaded areas):  The Zeeman field from Eq.~\eqref{eq:chemical_potentials}, $h>0$, shifts the unfilled band down to negative energies.  As a result, we begin occupying quasiparticle levels between $k_{\mathrm{min}}$ and $k_{\mathrm{max}}$ in the zero-temperature ground state.  For this range of momenta, the scattering states in the closed channel are Pauli blocked.}
\label{fig:Excitation_Spectrum}
\end{center}
\end{figure}

We now diagonalize $\mathcal{H}^\prime$ by using the Bogoliubov transformation:
\begin{align} \label{eq:Bogoliubov_transformation}
\psi_\k=\left(\begin{array}{l}g_{\up\k}\\e_{\down-\k}^{\dagger}\\g_{\down\k}\\e_{\up-\k}^{\dagger}\end{array}\right)=\left(\begin{array}{cccc}u_{{\k}}&v_{{\k}}&0&0\\-v_{{\k}}&u_{{\k}}&0&0\\0&0&u^\prime_{{\k}}&v^\prime_{{\k}}\\0&0&-v^\prime_{{\k}}&u^\prime_{{\k}}\end{array}\right)\left(\begin{array}{l}\beta_{\up\k}\\\beta_{\down-\k}^{\dagger}\\\zeta_{\down\k}\\\zeta_{\up-\k}^{\dagger}\end{array}\right)\hspace{-0.15em},\tag{25}
\end{align}
where $\{\beta_{\sigma\k},\,\zeta_{\sigma\k}\,;\,\sigma\in\{\up,\,\down\}\}$ are the operators for (fer-mionic) quasiparticle excitations in both channels, while $\{u_{\k},\,v_{\k}\}$ are real coefficients satisfying $u^2_{\k}+v^2_{\k}=1$ (and similarly for the primed variables).  Eventually, this yields the new Hamiltonian,
\begin{align} \label{eq:mean-field_Hamiltonian_(B)}
\mathcal{H}^\prime=\,\,&\Omega_0+\sum_{\k}\bigl[E_{\k}^o\beta_{\up\k}^\dagger\beta_{\up\k}+E_{\k}^o\beta_{\down-\k}^{\dagger}\beta_{\down-\k}\nn\\&+\,\bigl(E_{\k}^c-h\bigr)\zeta_{\down\k}^\dagger\zeta_{\down\k}+\bigl(E_{\k}^c+h\bigr)\zeta_{\up-\k}^{\dagger}\zeta_{\up-\k}\bigr]\hspace{+0.10em}.\tag{26}
\end{align}
The expression above contains the open- and closed-channel excitation energies,
\begin{align} \label{eq:excitation_energies}
E_{\k}^o&=\sqrt{\xi_{\k}^2+\Delta_o^2}\hspace{+0.20em},\nn\\E_{\k}^c\pm h&=\sqrt{\bigl(\xi_{\k}+\delta/2\bigr)^2+\Delta_c^2}\pm h\hspace{+0.20em}.\tag{27}
\end{align}
In the absence of spin imbalance, the zero-temperature grand potential is
\begin{align} \label{eq:ground-state_energy_(A)}
\Omega_0=\hspace{+0.25em}&\bigl[U_+\bigl(\Delta_o^2+\Delta_c^2\bigr)-2U_-\Delta_o\Delta_c\bigr]/\bigl(U_-^2-U_+^2\bigr)\nn\\&\hspace{-0.10em}+\sum_{\k}\bigl(2\xi_{\k}+\delta/2-E_{\k}^o-E_{\k}^c\bigr)\hspace{+0.10em}.\tag{28}
\end{align}
In this case, the eigenenergies for excitations are simply $\{E_{\k}^o,\,E_{\k}^c\}$ and there are no unpaired quasiparticles in the ground state.  However, in the presence of the Zeeman field, we begin populating quasiparticle modes in the ground state once $|h|>\mathrm{min}_{\k}E_{\k}^c$.  This condition defines a region of momenta, between $k_{\mathrm{min}}=\left\{2m\left[\mu-\delta/2-\left(h^2-\Delta_c^2\right)^{1/2}\right]\right\}^{1/2}$ and $k_{\mathrm{max}}=\left\{2m\left[\mu-\delta/2+\left(h^2-\Delta_c^2\right)^{1/2}\right]\right\}^{1/2}$, where the pairing gap is removed and we build up an unpaired Fermi sea in the closed channel (see Fig.~\ref{fig:Excitation_Spectrum}).  Under these considerations, we write the grand potential for the spin-imbalanced system as
\begin{align} \label{eq:ground-state_energy_(B)}
\Omega=\hspace{+0.10em}\Omega_0+\sum_{\k}\bigl[&\bigl(E_{\k}^c-h\bigr)\Theta\bigl(-E_{\k}^c+h\bigr)\bigr.\nn\\+\,&\bigl(E_{\k}^c+h\bigr)\Theta\bigl(-E_{\k}^c-h\bigr)\bigr]\hspace{+0.10em},\tag{29}
\end{align}
where the unit step function, $\Theta(\hspace{+0.14em}\cdots)$, selects out the range of states with negative quasiparticle energies.  (Since this formulation is symmetric with respect to a change in sign of the Zeeman field, we subsequently take $h>0$.)

\begin{figure}[t]
\begin{center}
\includegraphics[scale=0.35]{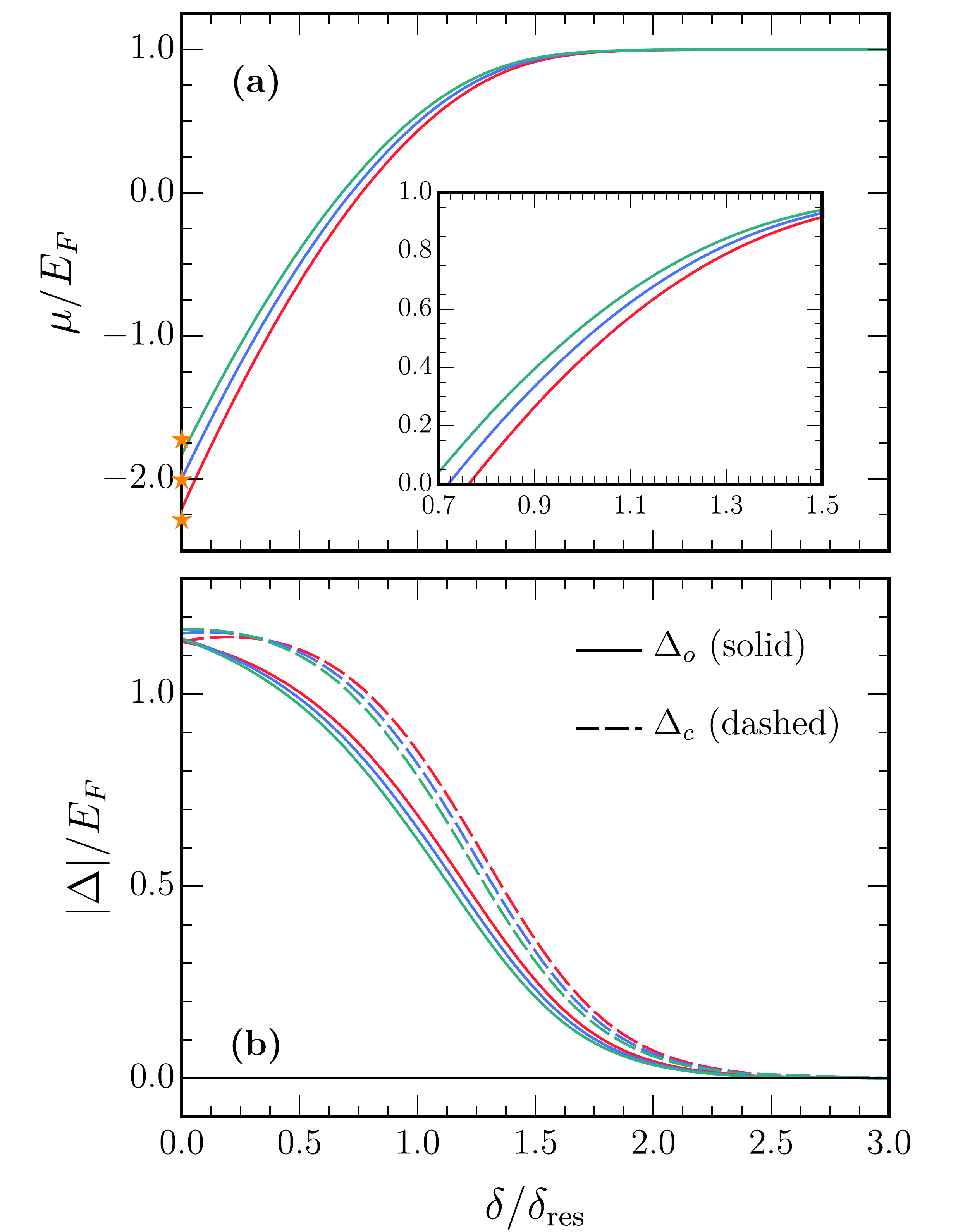}
\caption{The average chemical potential (a) and pairing order parameters (b) for the open and closed channels as functions of the interchannel detuning, $\delta/\delta_{\mathrm{res}}$, across the OFR.  The density of paired atoms is fixed for all plots at $n_{\mathrm{paired}}=10^{13}$ atoms per $\mathrm{cm}^{3}$, while the unpaired density in the closed channel is fixed to a different value for each plot:  $n_{\mathrm{unpaired}}=0$ (red, \protect\redlineB), $1\times10^{13}\,\mathrm{cm}^{-3}$ (blue, \protect\bluelineB), $2\times10^{13}\,\mathrm{cm}^{-3}$ (green, \protect\greenlineB).  To aid comparison, the Fermi energy, $E_F$, and Fermi momentum, $k_F$, are defined in terms of $n_{\mathrm{paired}}$.  Consequently, we have $\delta_\mathrm{res}/E_F\simeq15$ and $1/(k_Fa_T)\simeq1.5$ (where $a_T$ is the larger of the two fixed scattering lengths~\cite{Note4}).  In panel (a), the orange stars (\protect\starB) indicate the limiting behavior of the chemical potential on the BEC side (see Sec.~\ref{sec:BEC_Limit}), while a close-up view of the BCS side is provided in the inset.}
\label{fig:Many-Body_Plot_(A)}
\end{center}
\end{figure}

To study the crossover from BEC to BCS pairing, we minimize the grand potential in Eq.~\eqref{eq:ground-state_energy_(B)} with respect to both order parameters $\{\Delta_o,\,\Delta_c\}$ for given $\{\mu,\,h,\,\delta\}$.  By taking $\partial\Omega/\partial\Delta_o=0$ and $\partial\Omega/\partial\Delta_c=0$ to determine the stationary points, we obtain
\begin{align} \label{eq:gap_equations_(A)}
\Delta_o+\Delta_c&=-\frac{U_S}{2}\sum_{\k}\left\{\frac{\Delta_o}{E_{\k}^o}+\frac{\Delta_c}{E_{\k}^c}\bigl[1-\Theta\bigl(-E_{\k}^c+h\bigr)\bigr]\right\}\hspace{-0.10em},\nn\\\Delta_o-\Delta_c&=-\frac{U_T}{2}\sum_{\k}\left\{\frac{\Delta_o}{E_{\k}^o}-\frac{\Delta_c}{E_{\k}^c}\bigl[1-\Theta\bigl(-E_{\k}^c+h\bigr)\bigr]\right\}\hspace{-0.10em}.\tag{30}
\end{align}
Continuing, we renormalize the interactions in accordance with Eq.~\eqref{eq:renormalization}, and finally, we arrive at the two coupled gap equations:
\begin{align} \label{eq:gap_equations_(B)}
\Delta_o&=\frac{\Delta_c}{\chi_-}\left\{\frac{4\pi}{m}\sum_{\k}\left[\frac{1}{\epsilon_{\k}}-\frac{1-\Theta\bigl(-E_{\k}^c+h\bigr)}{E_{\k}^c}\right]-\chi_+\right\}\hspace{-0.10em},\nn\\\Delta_c&=\frac{\Delta_o}{\chi_-}\left\{\frac{4\pi}{m}\sum_{\k}\left[\frac{1}{\epsilon_{\k}}-\frac{1}{E_{\k}^o}\right]-\chi_+\right\}\hspace{-0.10em},\tag{31}
\end{align}
with $\chi_\pm=a_S^{-1}\pm a_T^{-1}$~\cite{Note4}.

We iteratively solve Eq.~\eqref{eq:gap_equations_(B)} and find two solutions:  one corresponds to a very deep global minimum of $\Omega$ where $\Delta_o$ and $\Delta_c$ are in phase, and the other is an excited saddlepoint with the gaps out of phase by $\pi$.  The former describes the case when all the particles are in a deep bound state created by the small $a_S$ in $^{173}$Yb (see Sec.~\ref{sec:Orbital_Feshbach_Resonance}).  Since the chemical potential for this state is large and negative at all values of the detuning~\cite{,PhysRevA.94.043624}, it is not related to the BCS-BEC crossover near the OFR.  Instead, the crossover physics is attributed to the shallow out-of-phase result which is well described at the low-energy scale of our model.  We can choose this particular solution by initially setting $U_S=0$ in Eq.~\eqref{eq:gap_equations_(A)} so that we have $\Delta_o=-\Delta_c$;  then we force the system to adiabatically maintain this phase difference for small $a_S$.

Along with the gap equations, we derive two separate number equations from Eq.~\eqref{eq:ground-state_energy_(B)}.  The average chemical potential $\mu$ is associated with the total number density, $n=n_o+n_c=-(\partial\Omega/\partial\mu)$, from which we obtain the densities in the open and closed channels,
\begin{align} \label{eq:number_densities}
n_o&=n_{g{\up}}+n_{e{\down}}=2n_{g{\up}}=\sum_{\k}\left\{1-\frac{\xi_{\k}}{E_{\k}^o}\right\}\hspace{-0.10em},\nn\\n_c&=n_{g{\down}}+n_{e{\up}}\nn\\&=\sum_{\k}\left\{1-\frac{\xi_{\k}+\delta/2}{E_{\k}^c}\bigl[1-\Theta\bigl(-E_{\k}^c+h\bigr)\bigr]\right\}\hspace{-0.10em}.\tag{32}
\end{align}
Similarly, the extra field $h$ corresponds to the density difference in the closed channel,
\begin{align} \label{eq:density_difference}
\delta n_c=n_{g{\down}}-n_{e{\up}}=-\frac{\partial\Omega}{\partial h}=\sum_{\k}\Theta\bigl(-E_{\k}^c+h\bigr)\hspace{+0.10em}.\tag{33}
\end{align}

\subsection{Results}
\label{sec:Results}

We now investigate how the BCS-BEC crossover physics close to the OFR is modified as a result of the closed-channel component in our model.  To properly isolate the effect of Pauli blocking, we fix the number of atoms that are involved in pairing,
\begin{align} \label{eq:n_paired}
n_{\mathrm{paired}}&=n_o+n_c-\delta n_c\nn\\&=n_{g{\up}}+n_{e{\down}}+2n_{e{\up}}\hspace{+0.15em},\tag{34}
\end{align}
and we consider different fixed values of $n_{\mathrm{unpaired}}/n_{\mathrm{paired}}$, where $n_{\mathrm{unpaired}}=\delta n_c$ is the number of unpaired atoms in the closed channel.  We also define the Fermi energy, $E_F=k_F^2/(2m)$, and the Fermi momentum, $k_F=\bigl(3\pi^2n_{\mathrm{paired}}\bigr)^{1/3}$, with respect to $n_{\mathrm{paired}}$ because this allows a comparison with the spin-balanced system.

\begin{figure}[ht!]
\begin{center}
\includegraphics[scale=0.42]{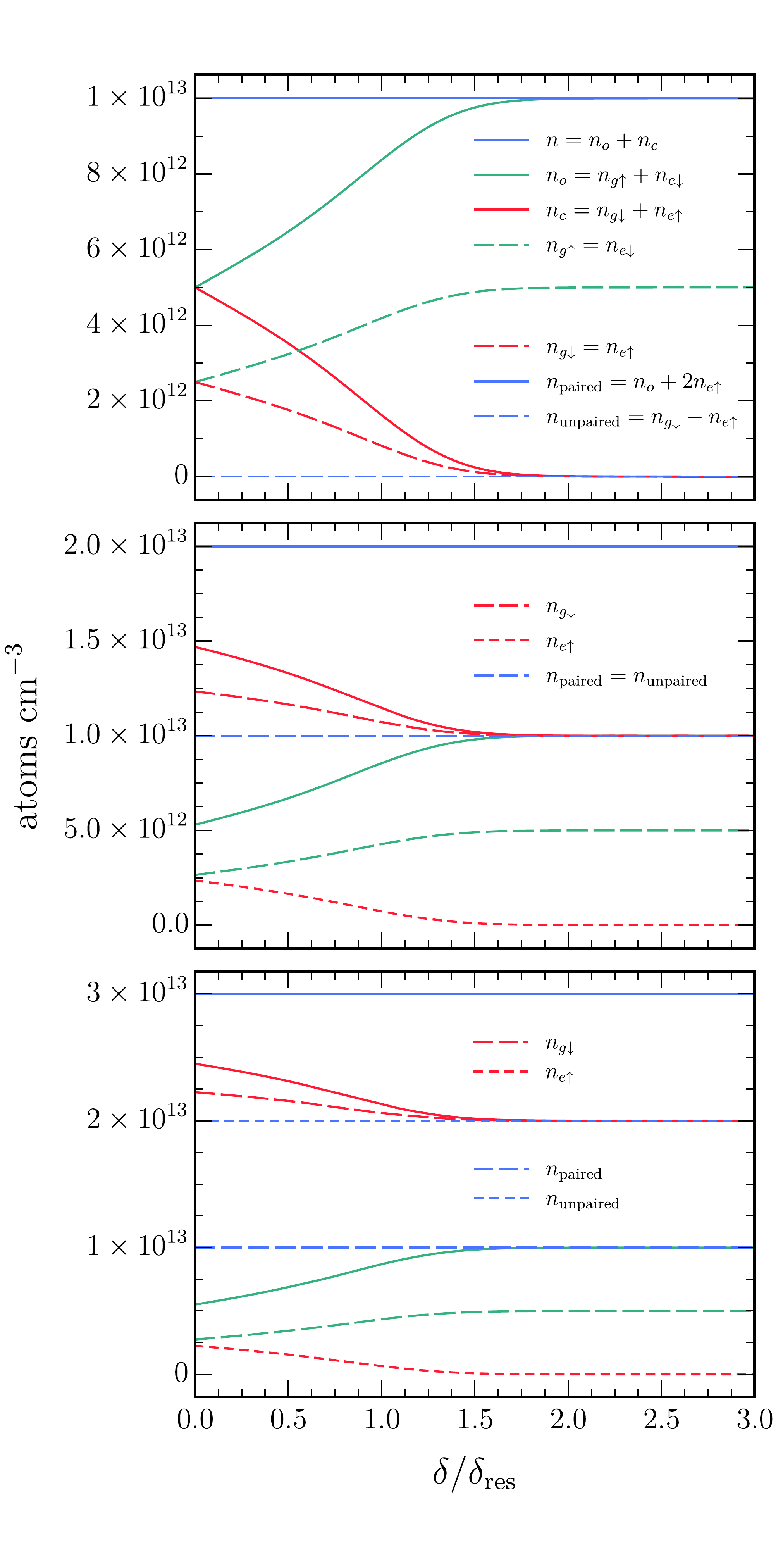}
\caption{Atom populations for the three crossover scenarios exhibited in Fig.~\ref{fig:Many-Body_Plot_(A)}:  $n_{\mathrm{unpaired}}/n_{\mathrm{paired}}=0$ (upper panel), 1 (middle panel), 2 (lower panel).  The total density, $n$ (blue solid lines), as well as the densities of paired and unpaired atoms, $n_{\mathrm{paired}}$ and $n_{\mathrm{unpaired}}$, are held fixed in each plot, and $n_{\mathrm{paired}}$ has the same value for all three plots.  While the open channel always possesses equal spin populations, $n_{g{\up}}=n_{e{\down}}$ (green dashed lines), the closed channel is characterized by a different imbalance, $n_{g{\down}}-n_{e{\up}}$, in each panel leading to the shifts in Fig.~\ref{fig:Many-Body_Plot_(A)}.}
\label{fig:Many-Body_Plot_(B)}
\end{center}
\end{figure}

In Fig.~\ref{fig:Many-Body_Plot_(A)}, we plot the crossover for $n_{\mathrm{paired}}=1\times10^{13}$ atoms per $\mathrm{cm}^{3}$ and with $n_{\mathrm{unpaired}}/n_{\mathrm{paired}}=0$ (red, \protect\redlineA), 1 (blue, \protect\bluelineA), 2 (green, \protect\greenlineA).  These densities are comparable to those achieved in current experiments~\cite{PhysRevLett.115.265302,PhysRevLett.122.193604}.  The upper panel (a) depicts the average chemical potential, $\mu/E_F$, against the interchannel detuning, $\delta/\delta_{\mathrm{res}}$, while the lower panel (b) shows the corresponding behaviors of the open- and closed-channel gaps, $|\Delta_o|/E_F$ and $|\Delta_c|/E_F$.  The red lines ($n_{g{\down}}=n_{e{\up}}$) resemble those in Fig.~3 of Ref.~\cite{PhysRevLett.115.135301}, although we have used updated values for the scattering lengths~\cite{Note4} and have also chosen a different paired-atom density.  Notably, here, in the BEC limit of zero detuning, we find that the chemical potential rises by $\sim0.2\,E_F$ when the unpaired-atom density increases by $n_{\mathrm{paired}}$.  Furthermore, at most magnetic fields, we see that $\mu/E_F$ is visibly shifted upwards, and $\{|\Delta_o|,\,|\Delta_c|\}/E_F$ downwards, for higher $n_{\mathrm{unpaired}}$.  This demonstrates how, by introducing a closed-channel medium that frustrates the interactions, we can effectively tune the crossover to suppress pairing.

For additional clarity, in Fig.~\ref{fig:Many-Body_Plot_(B)}, we show how the atom populations evolve as we sweep the magnetic field through the range in Fig.~\ref{fig:Many-Body_Plot_(A)}, with $n_{\mathrm{unpaired}}/n_{\mathrm{paired}}=0$ (upper panel), 1 (middle panel), 2 (lower panel).  Since we have a fixed total density $n$ in each panel, we observe that increasing the detuning drives atoms from the closed channel into the open channel.  Consequently, for large $\delta/\delta_{\mathrm{res}}$, the paired density moves fully into the open channel, $n_o\to n_{\mathrm{paired}}$, while the closed channel becomes either empty or completely polarized, $n_c\to n_{\mathrm{unpaired}}$.  Both channels in the upper plot have equal spin populations:  $n_{g{\up}}=n_{e{\down}}$ and $n_{g{\down}}=n_{e{\up}}$.  For the two remaining plots, we tune the Zeeman parameter $h/E_F$ while increasing $\delta/\delta_{\mathrm{res}}$, so that $n_{\mathrm{unpaired}}$ in the closed channel is fixed to a particular multiple of $n_{\mathrm{paired}}$.

\subsection{BEC Limit}
\label{sec:BEC_Limit}

For strong pairing, we can understand the shifts of the chemical potential in Fig.~\ref{fig:Many-Body_Plot_(A)} if we treat them as due to an effective interaction between open-channel pairs (or ``bosons'') and excess closed-channel atoms (fermions).  In the BEC regime where $\delta/\delta_{\mathrm{res}}\to0$ and $\mu/E_F\to-\infty$, we assume that the magnitude of the chemical potential in panel (a) is much greater than the gaps (b).  Therefore, we linearize Eq.~\eqref{eq:gap_equations_(B)} in the variables $\{\Delta_o,\,\Delta_c\}$, which affords the following implicit function for $\mu$:
\begin{widetext}
\begin{align} \label{eq:BEC_limit_(A)}
\def\arraystretch{2.8}\dfrac{\chi_-}{2\sqrt{2m|\mu|}-\chi_+}=\left\{\begin{array}{ll}\dfrac{2\sqrt{2m|\mu|}-\chi_+}{\chi_-}\hspace{+0.15em},\hspace{+0.10em}&|\mu|\geq h\hspace{+0.05em},\\\dfrac{2\sqrt{2m|\mu|}-\chi_+}{\chi_-}+\dfrac{2\sqrt{2}}{\pi\chi_-}\left[2\sqrt{m(h-|\mu|)}-2\sqrt{m|\mu|}\,\mathrm{arctan}\left(\sqrt{\dfrac{h}{|\mu|}-1}\,\right)\right]\hspace{-0.15em},\hspace{+0.10em}&|\mu|<h\hspace{+0.05em},\end{array}\right.\tag{35}
\end{align}
\end{widetext}
where $\chi_\pm$ are defined below Eq.~\eqref{eq:gap_equations_(B)}.  For the spin-balanced system ($\delta n_c=0$), the expression above has two solutions:  $\mu=-1/\bigl(2ma_S^2\bigr)$ and $\mu=-1/\bigl(2ma_T^2\bigr)$.  Both of these correspond to binding energies $\varepsilon_b$ where $\mu=-\varepsilon_b/2$, as required in the deep BEC limit, but only the latter is associated with the crossover (see Sec.~\ref{sec:Orbital_Feshbach_Resonance}).

Since the density of bosons is fixed across all three cases in Fig.~\ref{fig:Many-Body_Plot_(A)}, we expand the chemical potential around $-\varepsilon_b/2$ in powers of the excess density of fermions, $\delta n_c$.  To this end, we rewrite the right-hand side of Eq.~\eqref{eq:BEC_limit_(A)} as a Taylor series in $h-|\mu|$, and then we use Eq.~\eqref{eq:density_difference} to replace that small parameter by the excess density.  To first order this yields
\begin{align} \label{eq:BEC_limit_(B)}
\frac{\chi_-}{2\sqrt{2m|\mu|}-\chi_+}=\frac{2\sqrt{2m|\mu|}-\chi_+}{\chi_-}+\frac{4\pi\delta n_c}{m|\mu|\chi_-}\hspace{+0.15em},
\tag{36}
\end{align}
and solving for the chemical potential we then obtain $\mu\simeq-\varepsilon_b/2+g_{\mathrm{BF}}\delta n_c$.  Here, the binding energy is $\varepsilon_b=1/\bigl(ma_T^2\bigr)$ and $g_{\mathrm{BF}}=2\pi a_T/m$ is the effective coupling constant for the boson-fermion interaction in the Born approximation.  These first-order shifts are marked by the orange stars (\starA) in panel (a) and well capture the chemical potential behavior for BEC pairing.

\section{Conclusions and Outlook}
\label{sec:Conclusions_and_Outlook}

We have theoretically studied the orbital Feshbach resonance in the experimentally accessible $^{173}$Yb system.  Here, the closed channel is available as a set of well resolved intermediate states in the two-atom scattering process.  This is in contrast to more familiar alkali resonances, where the closed channel is detuned by an energy much larger than all other scales in the problem.  Following from this distinction, we investigated the effect of Pauli-blocking (frustration) caused by a closed-channel medium at both the few- and many-body levels.  By harmonically confining three particles to a single lattice site (with one atom always in the closed channel), we found that the lower lying eigenstates are strongly shifted relative to the two-body energy spectrum.  In the many-particle continuum picture, we saw that the introduction of a Fermi sea to the closed channel drives the system towards the BCS limit.

Our three-atom prediction could potentially be realized via a clock-spectroscopy scheme very similar to the recent experiment on two-dimensional polarons~\cite{PhysRevLett.122.193604}.  Although this experiment attempted the first observation of frustration, the largest Fermi seas that could be achieved, in practice, only led to shifts that were within the experimental uncertainties.  By contrast, the magnitude of our energy shifts suggests that an observation might be more feasible in the few-body scenario (on a lattice) than in the free-space system considered by that work.  For this reason, we furthermore expect that our many-body result would be enhanced in the context of a state-dependent optical lattice~\cite{PhysRevLett.120.143601}.  The findings of the current study may shed light on BCS pairing in more complicated solid-state multiband materials.  Additionally, our model of a Fermi sea in the presence of OFR-induced tunable pairing interactions may assist in the search for exotic superfluids, such as the sought after breached-pair phase~\cite{PhysRevLett.90.047002}.


\acknowledgments  We are grateful to N. Darkwah Oppong for helpful discussions on the experimental implementation of our three-body results.  We acknowledge support from the Australian Research Council (ARC) Centre of Excellence in Future Low-Energy Electronics Technologies (CE170100039).  Z.Y.S., M.M.P. and J.L. acknowledge financial support from the ARC via Discovery Project No.~DP160102739.  J.L. is additionally supported through the ARC Future Fellowship FT160100244.


\onecolumngrid

\appendix

\section{\hspace{1em}Derivation of the Three-Body Equations}
\label{sec:Appendix_A}

Here, we work through the method of solving the three-body problem discussed in Sec.~\ref{sec:Three-Body_Problem}.  To parameterize the motion of three equal-mass particles we define two relative coordinates, $\vect{x_{ij}}=\sqrt{1/2}\,(\vect{x}_j-\vect{x}_i)$ and $\vect{x_{k}^{ij}}=\sqrt{2/3}\,[(\vect{x}_i+\vect{x}_j)/2-\vect{x}_k]$, respectively, between atoms $i$ and $j$, and between their center of mass and atom $k$.  The square-root factors ensure the effective masses for the relative motions are both equal to the atomic mass $m$, and throughout this appendix we set $m=\omega=1$.  The coordinates, $\vect{x_{ij}}$ and $\vect{x_{k}^{ij}}$, are assigned sets of 3D harmonic-oscillator indices, $\vect{n_{ij}}\equiv\{\rho_{ij},\,l_{ij},\,\varsigma_{ij}\}$ and $\vect{n_{k}^{ij}}\equiv\{\rho_{k}^{ij},\,l_{k}^{ij},\,\varsigma_{k}^{ij}\}$, as done in the main text.  Now, the relative-motion equation decouples from the motion of the three-body center of mass with coordinate $\vect{X}=(\vect{x}_i+\vect{x}_j+\vect{x}_k)/3$.  Without loss of generality, therefore, we set the corresponding center-of-mass quantum numbers to zero, $\vect{N}=\{0,\,0,\,0\}$.  Since interactions are only in the atom-atom $s$-wave channel, we furthermore have $l_{ij}=\varsigma_{ij}=0$.  The atom-pair angular-momentum channels thus decouple, and we consider only the atom-pair $s$-wave states ($l_k^{ij}=\varsigma_k^{ij}=0$) as these are accessible in experiments.  This leads to a vanishing total angular momentum ($L+l_k^{ij}+l_{ij}=0$) and, as explained below, we consequently realize a pair of coupled equations depending only on a single index, $\rho_{k}^{ij}$.  [Note that $\{\rho_{ij},\,\rho_{k}^{ij}\}$ appear respectively as $\{\kappa,\,\rho\}$ in Eq.~\eqref{eq:matrix_kernels}.]

To begin, we make use of the three-body wave function, $|\Psi\rangle$ in Eq.~\eqref{eq:wave_function}, in addition to the noninteracting and interacting parts of the Hamiltonian, $H_0$ and $V$ in Eqs.~\eqref{eq:noninteracting_Hamiltonian} and~\eqref{eq:interaction_operator}-\eqref{eq:singlet_and_triplet_states}.  It is convenient to write $V$ in terms of the Huang-Yang pseudopotential~\cite{PhysRevA.93.042708}, $4\pi a_{S,\,T}\delta_{\mathrm{reg}}^{(3)}(\x_i-\x_j)$, where the singlet ($S$) and triplet ($T$) scattering lengths are $\{a_S,\,a_T\}$~\cite{Note4} and $\delta_{\mathrm{reg}}^{(3)}(\x)\equiv\delta^{(3)}(\vect{x})\del/\del x(x\,\cdot\,)$.  Then we project the Schr\"{o}dinger equation, $(E-H_0)|\Psi\rangle=V|\Psi\rangle$, separately onto both basis states of $|\Psi\rangle$.  Following projection onto $|g_{{\down}\vect{n_1}},e_{{\down}\vect{n_2}},g_{{\up}\vect{n_3}}\rangle$, we obtain
\begin{align} \label{eq:lambda_equation}
(E-E_{\vect{n}})\lambda_{\vect{n_1}\vect{n_2}\vect{n_3}}=4\pi\sum_{\vect{n_2'},\,\vect{n_3'}}\langle\vect{n_2},\vect{n_3}|\delta_{\mathrm{reg}}^{(3)}(\vect{x_2}-\vect{x_3})|\vect{n_2'},\vect{n_3'}\rangle\left[\frac12(a_S+a_T)\lambda_{\vect{n_1}\vect{n_2'}\vect{n_3'}}+\frac12(a_S-a_T)\gamma_{\vect{n_1}\vect{n_2'}\vect{n_3'}}\right]\hspace{-0.15em},\tag{A1}
\end{align}
while after projection onto $|g_{{\down}\vect{n_1}},e_{{\up}\vect{n_2}},g_{{\down}\vect{n_3}}\rangle$, we have
\begin{align} \label{eq:gamma_equation}
(E-E_{\vect{n}}-\delta)\gamma_{\vect{n_1}\vect{n_2}\vect{n_3}}&=4\pi\sum_{\vect{n_2'},\,\vect{n_3'}}\langle\vect{n_2},\vect{n_3}|\delta_{\mathrm{reg}}^{(3)}(\vect{x_2}-\vect{x_3})|\vect{n_2'},\vect{n_3'}\rangle\left[\frac12(a_S+a_T)\gamma_{\vect{n_1}\vect{n_2'}\vect{n_3'}}+\frac12(a_S-a_T)\lambda_{\vect{n_1}\vect{n_2'}\vect{n_3'}}\right]\nn\\&\hspace{+0.11em}-4\pi\sum_{\vect{n_1'},\,\vect{n_2'}}\langle\vect{n_1},\vect{n_2}|\delta_{\mathrm{reg}}^{(3)}(\vect{x_1}-\vect{x_2})|\vect{n_1'},\vect{n_2'}\rangle\left[\frac12(a_S+a_T)\gamma_{\vect{n_3}\vect{n_2'}\vect{n_1'}}+\frac12(a_S-a_T)\lambda_{\vect{n_3}\vect{n_2'}\vect{n_1'}}\right]\hspace{-0.15em}.\tag{A2}
\end{align}
Above, the state $|g_{{\down}\vect{n_1}},e_{{\down}\vect{n_2}},g_{{\up}\vect{n_3}}\rangle$ is defined as the zero of detuning and the single-particle energies of Eq.~\eqref{eq:noninteracting_Hamiltonian} are replaced by the notation, $E_{\vect{n}}\equiv\sum_{i\,=\,1}^3\varepsilon_{\vect{n_{i}}}$.

To reduce the number of summation indices we follow the general methods introduced in our earlier works~\cite{PhysRevA.96.032701,PhysRevA.97.042711}.  The idea is first to transform into the relative-motion frame and then to take the $\delta$-function boundary condition on the wave function that the colliding particles, $i$ and $j$, are superimposed on one another.  To this end, we insert a complete set of transformed states, $|\vect{n_1},\vect{n_2},\vect{n_3}\rangle\to|\vect{n}_{ij},\vect{n}_k^{ij},\vect{N}\rangle$, on either side of the pseudopotential in Eqs.~\eqref{eq:lambda_equation}-\eqref{eq:gamma_equation}.  Since the interaction only changes the quantum number for relative atom-atom motion $\vect{n}_{ij}$, this yields
\begin{subequations}
\begin{align}
\label{eq:relative_lambda}&(E-E_{\vect{n}})\lambda_{\vect{n_1}\vect{n_2}\vect{n_3}}\nn\\&=\sqrt{2}\pi\sum_{\widetilde{\vect{n}}',\,\vect{n_{23}'},\,\vect{n_{23}},\,\vect{n_1^{23}}}\langle\widetilde{\vect{n}}|\vect{n_{23}},\vect{n_1^{23}},0\rangle \phi_{\rho_{23}}\phi_{\rho_{23}'}\langle\vect{n_{23}'},\vect{n_1^{23}},0|\widetilde{\vect{n}}'\rangle\left[\frac12(a_S+a_T)\lambda_{\vect{n_1'}\vect{n_2'}\vect{n_3'}}+\frac12(a_S-a_T)\gamma_{\vect{n_1'}\vect{n_2'}\vect{n_3'}}\right]\hspace{-0.15em},\tag{A3}\\\label{eq:relative_gamma}&(E-E_{\vect{n}}-\delta)\gamma_{\vect{n_1}\vect{n_2}\vect{n_3}}\nn\\&=\sqrt{2}\pi\sum_{\widetilde{\vect{n}}',\,\vect{n_{23}'},\,\vect{n_{23}},\,\vect{n_1^{23}}}\langle\widetilde{\vect{n}}|\vect{n_{23}},\vect{n_1^{23}},0\rangle \phi_{\rho_{23}}\phi_{\rho_{23}'}\langle\vect{n_{23}'},\vect{n_1^{23}},0|\widetilde{\vect{n}}'\rangle\left[\frac12(a_S+a_T)\gamma_{\vect{n_1'}\vect{n_2'}\vect{n_3'}}+\frac12(a_S-a_T)\lambda_{\vect{n_1'}\vect{n_2'}\vect{n_3'}}\right]\nn\\&\hspace{+0.11em}-\sqrt{2}\pi\sum_{\widetilde{\vect{n}}',\,\vect{n_{21}'},\,\vect{n_{21}},\,\vect{n_3^{21}}}\langle\widetilde{\vect{n}}|\vect{n_{21}},\vect{n_3^{21}},0\rangle \phi_{\rho_{21}}\phi_{\rho_{21}'}\langle\vect{n_{21}'},\vect{n_3^{21}},0|\widetilde{\vect{n}}'\rangle\left[\frac12(a_S+a_T)\gamma_{\vect{n_3'}\vect{n_2'}\vect{n_1'}}+\frac12(a_S-a_T)\lambda_{\vect{n_3'}\vect{n_2'}\vect{n_1'}}\right]\hspace{-0.15em},\tag{A4}
\end{align}
\end{subequations}
where $\widetilde{\vect{n}}\equiv\{\vect{n_1},\,\vect{n_2},\,\vect{n_3}\}$ (and similarly for the primed variables).  Here, $\phi_{\rho_{ij}}=(1/\sqrt{4\pi}\,) \sqrt{2\rho_{ij}!/(\rho_{ij}+1/2)!}\,L_{\rho_{ij}}^{(1/2)}(0)$ are the $s$-wave eigenfunctions for the harmonic oscillator at zero relative separation, and $L$ is the associated Laguerre polynomial.

Next, we reformulate these expressions in terms of reduced atom-pair wave functions that have been subjected to the boundary condition.  We define two independent functions,
\begin{align} \label{eq:reduced_WFs(a)}
\eta_{\sigma\,=\,\lambda,\,\gamma}^{\vect{n_1^{23}}}=\sqrt{2}\pi\sum_{\widetilde{\vect{n}},\,\vect{n_{23}}}\phi_{\rho_{23}}\langle\vect{n_{23}},\vect{n_1^{23}},0|\widetilde{\vect{n}}\rangle\sigma_{\vect{n_1}\vect{n_2}\vect{n_3}}\hspace{+0.15em},\tag{A5}
\end{align}
by taking atoms 2 and 3 on top of each other.  By permuting the particle labels, we also have
\begin{align} \label{eq:reduced_WFs(b)}
\eta_{\sigma\,=\,\lambda,\,\gamma}^{\vect{n_3^{21}}}=\sqrt{2}\pi\sum_{\widetilde{\vect{n}},\,\vect{n_{21}}}\phi_{\rho_{21}}\langle\vect{n_{21}},\vect{n_3^{21}},0|\widetilde{\vect{n}}\rangle\sigma_{\vect{n_3}\vect{n_2}\vect{n_1}}\hspace{+0.15em},\tag{A6}
\end{align}
which correspond to when the boundary condition is applied to atoms 1 and 2.  After a direct substitution, Eqs.~\eqref{eq:relative_lambda}-\eqref{eq:relative_gamma} become
\begin{subequations}
\begin{align}
\label{eq:substituted_lambda}(E-E_{\vect{n}})\lambda_{\vect{n_1}\vect{n_2}\vect{n_3}}&=\sum_{\vect{n_{23}},\,\vect{n_1^{23}}}\langle\widetilde{\vect{n}}|\vect{n_{23}},\vect{n_1^{23}},0\rangle \phi_{\rho_{23}}\left[\frac12(a_S+a_T)\eta_\lambda^{\vect{n_1^{23}}}+\frac12(a_S-a_T)\eta_\gamma^{\vect{n_1^{23}}}\right]\hspace{-0.15em},\tag{A7}\\\label{eq:substituted_gamma}(E-E_{\vect{n}}-\delta)\gamma_{\vect{n_1}\vect{n_2}\vect{n_3}}&=\sum_{\vect{n_{23}},\,\vect{n_1^{23}}}\langle\widetilde{\vect{n}}|\vect{n_{23}},\vect{n_1^{23}},0\rangle \phi_{\rho_{23}}\left[\frac12(a_S+a_T)\eta_\gamma^{\vect{n_1^{23}}}+\frac12(a_S-a_T)\eta_\lambda^{\vect{n_1^{23}}}\right]\nn\\&\hspace{+0.11em}-\sum_{\vect{n_{21}},\,\vect{n_3^{21}}}\langle\widetilde{\vect{n}}|\vect{n_{21}},\vect{n_3^{21}},0\rangle \phi_{\rho_{21}}\left[\frac12(a_S+a_T)\eta_\gamma^{\vect{n_3^{21}}}+\frac12(a_S-a_T)\eta_\lambda^{\vect{n_3^{21}}}\right]\hspace{-0.15em}.\tag{A8}
\end{align}
\end{subequations}

Continuing, we divide Eq.~\eqref{eq:substituted_lambda} by $E-E_\n$ and Eq.~\eqref{eq:substituted_gamma} by $E-E_\n-\delta$, and then we act on the left (respectively) with the operators
\begin{align} \label{eq:extra_operators}
&\sqrt{2}\pi\sum_{\widetilde{\vect{n}},\,\vect{n_{23}'}}\phi_{\rho_{23}'}\langle\vect{n_{23}'},\vect{n_1^{23}},0|\widetilde{\vect{n}}\rangle(\,\cdot\,)\hspace{+0.20em},\nn\\&\sqrt{2}\pi\sum_{\widetilde{\vect{n}},\,\vect{n_{21}'}}\phi_{\rho_{21}'}\langle\vect{n_{21}'},\vect{n_3^{21}},0|\widetilde{\vect{n}}\rangle(\,\cdot\,)\hspace{+0.20em}.\tag{A9}
\end{align}
We thus arrive at
\begin{subequations}
\begin{align}
\label{eq:eta_lambda}\eta_\lambda^{\vect{n_1^{23}}}&=\sqrt{2}\pi\sum_{\widetilde{\n},\,\vect{n_{23}'},\,\vect{n_{23}},\,\vect{n_1^{23}}}\phi_{\rho_{23}'}\langle\vect{n_{23}'},\vect{n_1^{23}},0|\widetilde{\vect{n}}\rangle\frac{1}{E-E_{\vect{n}}}\langle\widetilde{\vect{n}}|\vect{n_{23}},\vect{n_1^{23}},0\rangle \phi_{\rho_{23}}\left[\frac12(a_S+a_T)\eta_\lambda^{\vect{n_1^{23}}}+\frac12(a_S-a_T)\eta_\gamma^{\vect{n_1^{23}}}\right]\hspace{-0.15em},\tag{A10}\\\label{eq:eta_gamma}\eta_\gamma^{\vect{n_3^{21}}}&=\sqrt{2}\pi\sum_{\widetilde{\n},\,\vect{n_{21}'}}\phi_{\rho_{21}'}\langle\vect{n_{21}'},\vect{n_3^{21}},0|\widetilde{\vect{n}}\rangle\frac{1}{E-E_{\vect{n}}-\delta}\left\{\sum_{\vect{n_{21}},\,\vect{n_3^{21}}}\langle\widetilde{\vect{n}}|\vect{n_{21}},\vect{n_3^{21}},0\rangle \phi_{\rho_{21}}\left[\frac12(a_S+a_T)\eta_\gamma^{\vect{n_3^{21}}}+\frac12(a_S-a_T)\eta_\lambda^{\vect{n_3^{21}}}\right]\right.\nn\\&\left.-\sum_{\vect{n_{23}},\,\vect{n_1^{23}}}\langle\widetilde{\vect{n}}|\vect{n_{23}},\vect{n_1^{23}},0\rangle \phi_{\rho_{23}}\left[\frac12(a_S+a_T)\eta_\gamma^{\vect{n_1^{23}}}+\frac12(a_S-a_T)\eta_\lambda^{\vect{n_1^{23}}}\right]\right\}\hspace{-0.10em}.\tag{A11}
\end{align}
\end{subequations}

After a series of straightforward manipulations~\cite{PhysRevA.96.032701,PhysRevA.97.042711}, and by recalling that the angular-momentum ($l,\,m$) quantum numbers are zero, we obtain a matrix equation in terms of the reduced wave functions:
\begin{align} \label{eq:eta_matrix}
\left(\begin{array}{c}\eta_\lambda\\\eta_\gamma\end{array}\right)=\frac12\left(\begin{array}{cc}(a_S+a_T)\mathbb{A}_E&(a_S-a_T)\mathbb{A}_E\\(a_S-a_T)(\mathbb{A}_{E-\delta}-\mathbb{B}_{E-\delta})&(a_S+a_T)(\mathbb{A}_{E-\delta}-\mathbb{B}_{E-\delta})\end{array}\right)\left(\begin{array}{c}\eta_\lambda\\\eta_\gamma\end{array}\right)\hspace{-0.10em}.\tag{A12}
\end{align}
Above, $\{\eta_\lambda,\,\eta_\gamma\}\equiv\{\eta_\lambda^{\rho},\,\eta_\gamma^{\rho}\}$ are vectors and $\{\mathbb{A}_E,\,\mathbb{B}_E\}\equiv\{\mathbb{A}_E^{\rho,\,\rho'},\,\mathbb{B}_E^{\rho,\,\rho'}\}$ are square matrices indexed by the radial atom-pair quantum number, $\rho\equiv\rho_{k}^{ij}$.  Equation~\eqref{eq:eta_matrix} may be solved for the energy $E$ as an eigenvalue problem which we rearrange into Eq.~\eqref{eq:matrix_equation} of the main text, with $\{\mathbb{A}_E,\,\mathbb{B}_E\}$ defined by Eq.~\eqref{eq:matrix_kernels}.  It only remains to be shown how the nondiagonal kernel $\mathbb{B}_E$ can be treated numerically, and the ensuing approach is inspired by Ref.~\cite{PhysRevA.82.023619}.

By using the notation of Eq.~\eqref{eq:matrix_kernels}, and now replacing $\{\x_{ij},\,\x_{k}^{ij}\}$ with $\{\x,\,\y\}$ for clarity, we write the kernel explicitly:
\begin{align} \label{eq:matrix_B(a)}
\mathbb{B}_E^{\rho,\,\rho'}=\frac{1}{32\sqrt{2}\pi^2}\sum_{\kappa,\,\kappa'}\frac{R_{\kappa}(0)R_{\kappa'}(0)}{E-2\kappa'-2\rho'}\int d^{(3)}\x\,d^{(3)}\y\,R_{\kappa}(|\x|)\,R_{\rho}(|\y|)\,R_{\kappa'}\left(\left|\frac{\x-\sqrt{3}\y}{2}\right|\right)R_{\rho'}\left(\left|\frac{\sqrt{3}\x+\y}{2}\right|\right)\hspace{-0.10em},\tag{A13}
\end{align}
where the zero-angular-momentum eigenfunctions for the harmonic oscillator are given by
\begin{align} \label{eq:HO_eigenfunctions}
\frac{1}{2\sqrt{\pi}}R_{\kappa}(|\x|\equiv x)=\frac{1}{2\sqrt{\pi}}\sqrt{\frac{2\kappa!}{(\kappa+1/2)!}}L_\kappa^{(1/2)}\bigl(x^2\bigr)\,\mathrm{exp}\left(-\frac{x^{2}}{2}\right)\hspace{-0.10em},\tag{A14}
\end{align}
with $R$ the radial component~\cite{SMIRNOV1962346}.

For each coordinate in Eq.~\eqref{eq:matrix_B(a)}, there is a Green's function $G$ which solves the noninteracting Schr\"{o}dinger equation with a Dirac-delta-function source.  As an example, we have (in first quantization)
\begin{align} \label{eq:Schrodinger_equation}
\left[\nu+\frac32-\left(-\frac{\nabla_\x^{2}}{2}+\frac{x^{2}}{2}\right)\right]G_{\nu}(\x)=\delta^{(3)}(\x)\tag{A15}
\end{align}
for the $\x$ coordinate, where $\nabla_\x^{2}$ is the 3D Laplacian operator and the energy $\nu$ is defined relative to the zero-point energy of $\frac32$.  By expanding both functions in the complete set of harmonic-oscillator eigenfunctions, we obtain
\begin{align} \label{eq:HO_expansions}
\delta^{(3)}(\x)=\frac{1}{4\pi}\sum_{\kappa}R_{\kappa}(0)R_{\kappa}(x)\quad\mathrm{and}\quad G_{\nu}(\x)=\frac{1}{4\pi}\sum_{\kappa}\frac{R_{\kappa}(0)R_{\kappa}(x)}{\nu-2\kappa}\hspace{+0.20em}.\tag{A16}
\end{align}
These relations allow us to convert Eq.~\eqref{eq:matrix_B(a)} into a one-dimensional integral, as shown:
\begin{align} \label{eq:matrix_B(b)}
\mathbb{B}_E^{\rho,\,\rho'}&=\frac{1}{8\sqrt{2}\pi}\sum_{\kappa}R_{\kappa}(0)\int d^{(3)}\x\,d^{(3)}\y\,G_{E-2\rho'}\left(\frac{\x-\sqrt{3}\y}{2}\right)R_{\kappa}(x)\,R_{\rho}(y)\,R_{\rho'}\left(\left|\frac{\sqrt{3}\x+\y}{2}\right|\right)\nn\\&=\frac{1}{2\sqrt{2}}\int d^{(3)}\x\,d^{(3)}\y\,\delta^{(3)}(\x)\,G_{E-2\rho'}\left(\frac{\x-\sqrt{3}\y}{2}\right)R_{\rho}(y)\,R_{\rho'}\left(\left|\frac{\sqrt{3}\x+\y}{2}\right|\right)\nn\\&=\sqrt{2}\pi\int_0^\infty dy\,y^2\,G_{E-2\rho'}\left(-\frac{\sqrt{3}\y}{2}\right)R_{\rho}(y)\,R_{\rho'}\left(\frac y2\right)\hspace{-0.05em},\tag{A17}
\end{align}
and this agrees with the corresponding result in Ref.~\cite{PhysRevA.82.023619}.

Now, to evaluate the Green's function, we simply replace $R_{\kappa}(0)$ in Eq.~\eqref{eq:HO_expansions} with its definition from Eq.~\eqref{eq:HO_eigenfunctions}.  Then the identities,
\begin{align} \label{eq:extra_identities}
L_\kappa^{(1/2)}(0)=\begin{pmatrix}\kappa+1/2\\\kappa\end{pmatrix}=\frac{(\kappa+1/2)!}{\kappa!\,(1/2)!}\quad\mathrm{and}\quad(1/2)!=\frac{\sqrt{\pi}}{2}\hspace{+0.20em},\tag{A18}
\end{align}
lead us to write down
\begin{align} \label{eq:Green's_function}
G_{\nu}(\x)&=\frac{1}{\sqrt{2\pi^3}}\sum_\kappa\sqrt{\frac{(\kappa+1/2)!}{\kappa!}}\frac{R_\kappa(x)}{\nu-2\kappa}\hspace{+0.20em}.\tag{A19}
\end{align}

Therefore, we finally arrive at the following equivalent form of the nondiagonal kernel in Eq.~\eqref{eq:matrix_B(a)}:
\begin{align} \label{eq:matrix_B(c1)}
\mathbb{B}_E^{\rho,\,\rho'}=\frac{1}{\sqrt{\pi}}\sum_\kappa\sqrt{\frac{(\kappa+1/2)!}{\kappa!}}\frac{N_{\kappa,\,\rho,\,\rho'}}{E-2(\kappa+\rho')}\tag{A20}
\end{align}
where
\begin{align} \label{eq:matrix_B(c2)}
N_{\kappa,\,\rho,\,\rho'}=\int_{0}^{\infty}dy\,y^2\,R_{\kappa}\left(\frac{\sqrt{3}y}{2}\right)R_{\rho}(y)\,R_{\rho'}\left(\frac y2\right)\hspace{-0.05em}.\tag{A21}
\end{align}
Reference~\cite{PhysRevA.82.023619} found that for a cutoff $\rho_{\mathrm{max}}>\{\rho,\,\rho'\}$, the integral $N_{\kappa,\,\rho,\,\rho'}$ vanishes for large enough $\kappa>\kappa_{\mathrm{max}}\sim4\rho_{\mathrm{max}}$.  Numerically, this means that we can tabulate the values of $N$ first and then populate the matrix $\mathbb{B}_E$ by a simple and fast summation over $\kappa$.  We can subsequently see that the kernel is symmetric, $\mathbb{B}_E^{\rho,\,\rho'}=\mathbb{B}_E^{\rho',\,\rho}$.

\twocolumngrid

\bibliography{OFR_Refs.bib}

\onecolumngrid

\end{document}